# REASONING ABOUT TRANSFINITE SEQUENCES*

Stéphane Demri

*LSV, CNRS & INRIA Futurs projet SECSI & ENS Cachan*
*61 avenue du Président Wilson, 94235 Cachan Cedex, France*
*Email:* `demri@lsv.ens-cachan.fr`

and

David Nowak

*Research Center for Information Security*
*National Institute of Advanced Industrial Science and Technology*
*Akihabara Daibiru, room 1102, 1-18-13 Sotokanda, Chiyoda-ku, Tokyo 101-0021, Japan*
*Email:* `nowak@yl.is.s.u-tokyo.ac.jp`

ABSTRACT

We introduce a family of temporal logics to specify the behavior of systems with Zeno behaviors. We extend linear-time temporal logic LTL to authorize models admitting Zeno sequences of actions and quantitative temporal operators indexed by ordinals replace the standard next-time and until future-time operators. Our aim is to control such systems by designing controllers that safely work on $\omega$-sequences but interact synchronously with the system in order to restrict their behaviors. We show that the satisfiability and model-checking for the logics working on $\omega^k$-sequences is EXPSPACE-complete when the integers are represented in binary, and PSPACE-complete with a unary representation. To do so, we substantially extend standard results about LTL by introducing a new class of succinct ordinal automata that can encode the interaction between the different quantitative temporal operators.

## 1. Introduction

**Control of physical systems.** Modelling interaction between a computer system and a physical system has to overcome the difficulty of the different time scales. For example, reasoning about the connection between the physical description of an electric circuit and its logical description in VHDL (standard language designed and optimized for describing the behavior of digital systems) needs to take into account that the two descriptions are dealing with objects running at distinct speeds. The speeds can be so different that some abstraction consists in assuming one system evolves infinitely quicker than the other one. Another kind of interaction consists of controlling a physical system by a computer system. Usually, a physical system is modelled by differential equations. Solving those equations can then involve computations of limits. For instance, in the bouncing ball example [20], in a finite amount of time an infinite number of actions can be performed. It is a Zeno sequence of actions. Similar behaviors have also been considered to solve the car-bee problem [26]. However, Zeno behaviors are usually excluded from the modelling of real-time controllers, which is a reasonable requirement (see e.g. [10]), but also from the modelling of the physical systems, see some exception in [8, 30]. This is a quite drastic limitation, since Zeno sequences are often acceptable behaviors for physical systems.

---

*The first author acknowledges partial support by the ACI "Sécurité et Informatique" COR-TOS. The second author acknowledges partial support by the e-Society project of MEXT. Part of this work was done while the second author was affiliated to LSV, CNRS & ENS de Cachan and Department of Information Science, The University of Tokyo.

**Beyond $\omega$-sequences.** Our main motivation in this paper is to model Zeno behaviors and ultimately to control physical systems admitting such behaviors. To do so, we introduce a specification logical language that is interpreted on well-ordered linear orderings. Reasoning problems based on this logical language should admit efficient algorithms, as good as those for standard specification languages as linear-time temporal logic LTL, see e.g. [19]. The $\omega$-sequences are already familiar objects in model-checking, see e.g. [45], even though such infinite objects are never manipulated when model-checking finite-state programs. Indeed, most problems on Büchi automata reduce to standard reachability questions on finite graphs. In a similar fashion, the behaviors of physical systems are modeled in the paper by sequences indexed by countable ordinals (see e.g. [41]), i.e. equivalence classes of well-ordered linear orderings, even though as we will show most problems will also reduce to questions on finite graphs. For instance, the law of movement of the bouncing ball is modelled by a set of sequences of length $\omega^2$. The specification of the ball, i.e. the set of acceptable behaviors, is also characterized as a set of sequences of the same length $\omega^2$. On the other hand, the controller is a computer system whose complete executions are $\omega$-sequences. In this paper, we allow Zeno behaviors of physical systems and we will present a specification language working on sequences indexed by ordinals greater than the usual first infinite ordinal $\omega$.

**Our contribution.** We introduce a class of logics LTL($\alpha$) indexed by a countable ordinal $\alpha$ closed under addition whose models are sequences of length $\alpha$. Quantitative extensions of the standard next-time X and until U operators are considered by allowing operators of the form $X^\beta$ and $U^\beta$ with $\beta$ smaller than $\alpha$. As shown in the paper, for every $\alpha \leq \omega^\omega$, LTL($\alpha$) can be viewed as a fragment of the monadic second-order theory $\langle \omega^\omega, < \rangle$ known to be decidable, see e.g. [13]. For every $k \geq 1$, we show that LTL($\omega^k$) satisfiability is PSPACE-complete with an unary encoding of integers and EXPSPACE-complete with a binary encoding. This generalizes non-trivially what is known about LTL. We reduce the satisfiability problem to the non-emptiness problem of ordinal automata recognizing transfinite words [12, 18, 46, 28, 11]. The reduction entails that the satisfiability problem has an elementary complexity (by using [16]) but does not guarantee the optimal upper bound. We introduce a class of succinct ordinal automata of level $k$, $k \geq 1$ in which the LTL($\omega^k$) formulae can be translated into and we prove that the non-emptiness problem is in NLOGSPACE. Succinctness allows us to reduce by one exponential the size of the automata obtained by translation which provides us the optimal upper bound. Analogous complexity results are shown for model checking. Finally, we introduce and motivate a control problem with inputs a physical system $\mathcal{S}$ modelled by an ordinal automaton working on $\omega^k$-sequences, and an LTL($\omega^k$) formula $\phi$ describing the desirable behaviors of the system. The problem we introduce is the existence of a controller $\mathcal{C}$ working on $\omega$-sequences such that all the behaviors of $\mathcal{S} \times \mathcal{C}$ satisfy the property $\phi$. The synchronization operation $\times$ takes into account the different time scales between $\mathcal{S}$ and $\mathcal{C}$ and the set of synchronization vectors depending on the set of observable actions of the controller $\mathcal{C}$. As a by-product of our results, checking whether a controller satisfies the above conditions can be done effectively but we leave the question of the existence and synthesis of such controllers for future work.

**Related work.** Our original motivation in this work is the control of systems with legal Zeno behaviors by systems whose complete executions are $\omega$-sequences. The theory of control of discrete event systems was introduced in [38]. In this theory, a process is a deterministic non-complete finite automaton over an alphabet of events. The control problem consists in, given a process $P$ and a set $S$ of admissible behaviors, finding a process $Q$ such that the behaviors of $P \times Q$ are in $S$ and such that $Q$ reacts to all uncontrollable events and cannot detect unobservable events. Extension to specifications from the modal $\mu$-calculus can be found in [3] whereas the control of timed systems (without Zeno behaviors) is for instance studied in [4, 29, 10]. It is plausible that the techniques from the above-mentioned works (see also [36, 43, 2]) can be adapted to the control problem we have introduced but the technical contribution of this paper is mainly oriented towards satisfiability and model-checking issues.

The logics we have introduced belong to the long tradition of quantitative versions of LTL. LTL-like logics having models non isomorphic to $\omega$ can be found in [1, 40, 39, 21, 32, 34]. Temporal operators in the real-time logics from [1, 32, 34] are indexed by intervals as our logics LTL($\alpha$). However, among the above-mentioned works, Rohde's thesis [40] contains a LTL-like logic interpreted over $\alpha$-sequences with ordinal $\alpha$ but the temporal operators are simply the standard next-time and until operators without any decoration. It is shown in [40] that the satisfiability problem for such a logic can be decided in exponential-time when the inputs are the formula to be tested and the countable ordinal from which the model is built. Similarly, in [5] a temporal logic with next-time and sometimes operators but interpreted over well-founded trees of $\omega$-segments is shown decidable by designing a cut-free sequent-style calculus. The concept of time gaps in [5] can be put naturally in correspondence with limits for ordinals. No complexity issues are discussed in [5] even though the temporal logic with only temporal operator next-time is shown equivalent to the famous modal logic S4 interpreted over reflexive and transitive Kripke frames and known to be PSPACE-complete.

In the paper, we follow the automata-based approach for temporal logics from [45] but we deal with ordinal automata recognizing words of length $\alpha$ for some countable ordinal $\alpha$. So, we extend the reduction from LTL into generalized Büchi automata to the reduction from LTL($\omega^k$) into ordinal automata recognizing words of length $\omega^k$. Many classes of ordinal automata have been introduced in the literature. In [12, 18] automata recognizing $\omega^k$-sequences for some $k \geq 1$ are introduced making essential the concept of layer. In [13, 46, 28], such automata are generalized to recognize $\alpha$-sequences for $\alpha$ countable. Correspondences between these different classes can be found in [7]. In the paper, we mainly adopt the definitions from [28]. An elegant and powerful extension to automata recognizing words indexed elements from a linear ordering can be found in [11]. As far as we know, automata recognizing sequences of length greater than $\omega$ designed to solve verification problems have been first used in [25] to model concurrency by limiting the state explosion problem. Similarly, timed automata accepting Zeno words are introduced in [8] in order to model physical phenomena with convergent executions. The non-emptiness problem for such automata is shown to be decidable [8].

As LTL can be viewed as the first-order fragment of monadic second order theory over $\langle \mathbb{N}, < \rangle$, theories over $\langle \alpha, < \rangle$ for some countable ordinal $\alpha$ have been also studied by Büchi [12], see also [13, 7]. For instance, decidability of monadic second order theories over $\langle \alpha, < \rangle$ for some countable ordinal $\alpha$ is shown in [13]. Decidability status of elementary theories over countable ordinals have been established in [9, 17].

**Plan of the paper.** In Sect. 2, we recall basic definitions about ordinals and we introduce a class of linear-time temporal logics parameterized by the length of the models. In Sect. 2.4, we show that any logic admitting models of length $\alpha$ with $\alpha \leq \omega^\omega$ is decidable by translation into the decidable monadic second order theory $\langle \omega^\omega, < \rangle$. Sect. 3 shows how the class of models of a given formula from a logic working on $\omega^k$-sequences ($k < \omega$) can be recognized by an ordinal automaton. To do so, we substantially extend what is known about LTL with generalized Büchi automata. In order to fully characterize the complexity of logics working on $\omega^k$-sequences (EXPSPACE-completeness or PSPACE-completeness depending on the way integers are encoded), in Sect. 4 we introduce a class of succinct ordinal automata of level $k$, extending generalized Büchi automata, and we show that the emptiness problem is NLOGSPACE-complete. In Sect. 5, since we have at this point all the necessary background, we present the control problem that motivates our investigations. We prove that we can decide whether a given controller satisfies the properties stated in our logical framework. Sect. 6 contains concluding remarks and open problems.

This paper is a completed version of [22]. Full proofs can be found in the technical

appendix

## 2. Temporal Logics on Transfinite Sequences

*2.1. Ordinals*

We recall basic definitions and properties about ordinals, see e.g. [41] for additional material. An *ordinal* is a totally ordered set which is *well ordered*, i.e. all its non-empty subset have a least element. Order-isomorphic ordinals are considered equal. They can be more conveniently defined inductively by: the empty set (written 0) is an ordinal, if $\alpha$ is an ordinal, then $\alpha \cup \{\alpha\}$ (written $\alpha + 1$) is an ordinal and, if $X$ is a set of ordinal, then $\bigcup_{\alpha \in X} \alpha$ is an ordinal. The ordering is obtained by $\beta < \alpha$ iff $\beta \in \alpha$. An ordinal $\alpha$ is a *successor* ordinal iff there exists an ordinal $\beta$ such that $\alpha = \beta + 1$. An ordinal which is not 0 or a successor ordinal, is a *limit* ordinal. The first limit ordinal is written $\omega$. Addition, multiplication and exponentiation can be defined on ordinals inductively: $\alpha + 0 = \alpha$, $\alpha + (\beta + 1) = (\alpha + \beta) + 1$ and $\alpha + \beta = sup\{\alpha + \gamma : \gamma < \beta\}$ where $\beta$ is a limit ordinal. Multiplication and exponentiation are defined similarly. $\epsilon_0$ is the closure of $\omega \cup \{\omega\}$ under ordinal addition, multiplication and exponentiation. By the Cantor Normal Form theorem, for any ordinal $\alpha < \epsilon_0$, there are unique ordinals $\beta_1, \ldots, \beta_p$, and unique integers $n_1, \ldots, n_p$ such that $\alpha > \beta_1 > \cdots > \beta_p$ and

$$\alpha = \omega^{\beta_1} \times n_1 + \cdots + \omega^{\beta_p} \times n_p$$

If $\alpha < \omega^\omega$, then the $\beta_i$'s are integers.

Whenever $\alpha \leq \beta$, there is a unique ordinal $\gamma$ such that $\alpha + \gamma = \beta$. We write $\beta - \alpha$ to denote $\gamma$. For instance, $\omega^2 - \omega = \omega^2$, $\omega \times 3 - \omega = \omega \times 2$ and $\omega^2 - \omega^3$ is not defined since $\omega^3 > \omega^2$.

Given an ordinal $\alpha \leq \omega^k$ equal to $\omega^k a_k + \cdots + \omega^1 a_1 + \omega^0 a_0$, we write $sum(\alpha)$ to denote $a_k + \cdots + a_0$, $head(\alpha)$ to denote the maximal $i$ such that $a_i \neq 0$ and $tail(\alpha)$ to denote the minimal $i$ such that $a_i \neq 0$ (assuming $\alpha \neq 0$). For instance, $tail(\alpha + \omega^n) = n$.

An ordinal $\alpha$ is said to be closed under addition whenever $\beta, \beta' < \alpha$ implies $\beta + \beta' < \alpha$. For instance, 0, 1, $\omega$, $\omega^2$, $\omega^3$, and $\omega^\omega$ are closed under addition. In the sequel, we shall consider logics whose models are $\alpha$-sequences, i.e. mappings of the form $\alpha \to \Sigma$ for some finite alphabet $\Sigma$ and ordinal $\alpha$ closed under addition.

**Lemma 1** *For every ordinal $\alpha \geq 1$, $\alpha$ is closed under addition iff its Cantor normal form is $\omega^\beta$ for some ordinal $\beta$.*

The proof of Lemma 1 can be found in Appendix A.

*2.2. Quantitative Extensions of LTL*

For every ordinal $\alpha$ closed under addition, we introduce the logic LTL($\alpha$) whose models are precisely sequences of the form $\sigma : \alpha \to 2^{\mathrm{AP}}$ for some countably infinite set AP of atomic propositions. The formulae of LTL($\alpha$) are defined as follows:

$$\phi ::= p \ | \ \neg \phi \ | \ \phi_1 \wedge \phi_2 \ | \ \mathtt{X}^\beta \phi \ | \ \phi_1 \mathtt{U}^{\beta'} \phi_2,$$

where $p \in \mathrm{AP}$, $\beta < \alpha$ and $\beta' \leq \alpha$. The satisfaction relation is inductively defined below where $\sigma$ is a model for $\mathrm{LTL}(\alpha)$ and $\beta < \alpha$:

- $\sigma, \beta \models p$ iff $p \in \sigma(\beta)$,
- $\sigma, \beta \models \neg \phi$ iff not $\sigma, \beta \models \phi$; $\sigma, \beta \models \phi_1 \wedge \phi_2$ iff $\sigma, \beta \models \phi_1$ and $\sigma, \beta \models \phi_2$,
- $\sigma, \beta \models \mathtt{X}^{\beta'} \phi$ iff $\sigma, \beta + \beta' \models \phi$,
- $\sigma, \beta \models \phi_1 \mathtt{U}^{\beta'} \phi_2$ iff there is $\gamma < \beta'$ such that $\sigma, \beta + \gamma \models \phi_2$ and for every $\gamma' < \gamma$, $\sigma, \beta + \gamma' \models \phi_1$.

Closure under addition of $\alpha$ guarantees that $\beta + \beta'$ and $\beta + \gamma$ above are strictly smaller than $\alpha$. Moreover, in $\mathtt{X}^{\beta'} \phi$, $\beta' \leq \alpha$ so that for any $\beta < \alpha$, $\beta + \beta' < \alpha$. By contrast, in $\phi_1 \mathtt{U}^{\beta'} \phi_2$, $\beta' \leq \alpha$ (not necessarily strictly) because satisfaction of $\sigma, \beta \models \phi_1 \mathtt{U}^{\beta'} \phi_2$ implies the existence of some $\gamma$ (satisfying some conditions) that is already strictly less than $\beta'$. The models of $\phi \in \mathrm{LTL}(\alpha)$ are defined as elements of the set $\mathrm{Mod}(\phi) = \{\sigma : \sigma, 0 \models \phi\}$. $\phi$ is said to be $\mathrm{LTL}(\alpha)$-satisfiable whenever $\mathrm{Mod}(\phi)$ is non-empty.

The operator $\mathtt{X}^\beta$ is a natural generalization of the next-time operator from linear-time temporal logic LTL that allows to perform a jump of fixed length $\beta$. Similarly, the operator $\mathtt{U}^\beta$ is a natural generalization of the until operator from LTL. We extend the standard abbreviations as follows: $\mathtt{F}^\beta \phi \stackrel{\mathrm{def}}{=} \top \mathtt{U}^\beta \phi$ and $\mathtt{G}^\beta \phi \stackrel{\mathrm{def}}{=} \neg \mathtt{F}^\beta \neg \phi$.

The logic $\mathrm{LTL}(1)$ is equivalent to the propositional calculus since $\phi_1 \mathtt{U}^0 \phi_2$ is equivalent to $\bot$, $\phi_1 \mathtt{U}^1 \phi_2$ is equivalent to $\phi_2$, and $\mathtt{X}^0 \phi$ is equivalent to $\phi$. LTL is expressively equivalent to $\mathrm{LTL}(\omega)$: the operators $\mathtt{X}^n$ and $\mathtt{U}^n$ for $n \geq 0$, and $\mathtt{U}^\omega$ can be simply expressed with the LTL operators $\mathtt{X}$ and $\mathtt{U}$. However, $\mathrm{LTL}(\omega)$ is more succinct than LTL if the natural numbers are encoded with a binary representation (see Lemma 8).

Actually in order to study the decidability/complexity of $\mathrm{LTL}(\alpha)$, we restrict ourselves to countable limit ordinals $\alpha$ so that the set of formulae is itself countable. Furthermore, for studying complexity issues, it is necessary to specify the encoding of the ordinals $\beta \leq \alpha$ occurring in $\mathrm{LTL}(\alpha)$ formulae. In the sequel, we use Cantor normal form to encode ordinals $1 \leq \beta \leq \omega^\omega$, and the natural numbers occurring in such normal forms are represented in binary.

We provide below properties dealing with limit states that can be easily expressed in $\mathrm{LTL}(\omega^k)$ ($k \geq 2$)

1. "$p$ holds in the states indexed by limit ordinals strictly less than $\omega^k$":
$$\mathtt{G}^{\omega^k}(\mathtt{X}^\omega p \wedge \cdots \wedge \mathtt{X}^{\omega^{k-1}} p).$$

2. For $1 \leq k' \leq k-2$, "if $p$ holds infinitely often in states indexed by ordinals of the form $\omega^{k'} \times n$, $n \geq 1$, then $q$ holds in the state indexed by $\omega^{k'+1}$":
$$(\mathtt{G}^{\omega^{k'+1}} \mathtt{F}^{\omega^{k'+1}} \mathtt{X}^{\omega^{k'}} p) \Rightarrow (\mathtt{X}^{\omega^{k'+1}} q).$$

*2.3. Model-checking*

The model-checking checking for LTL($\alpha$) is defined as a natural extension of the problem for LTL (its existential version) where the labelled transition systems are replaced by automata recognizing $\alpha$-sequences (see Definition 1).

*Model-checking problem for* LTL($\alpha$):

**input** : An ordinal automaton $\mathcal{A}$ with finite alphabet a subset of $2^{\mathrm{AP}}$ (see Sect. 3.1 for a definition) and an LTL($\alpha$) formula $\phi$.

**question:** Is there an $\alpha$-sequence $\sigma$ accepted by $\mathcal{A}$ such that $\sigma, 0 \models \phi$?

This is the existential version of model-checking (easier to relate with the satisfiability problem). The universal variant of the problem asks whether for all the $\alpha$-sequences $\sigma$ accepted by $\mathcal{A}$, we have $\sigma, 0 \models \phi$.

By standard arguments in computational complexity about deterministic classes and since LTL($\alpha$) is closed under negation, the complexity results for the existential variant of model checking about PSPACE-completeness and EXPSPACE-completeness, holds also true for the universal variant. Moreover, it is worth observing that in ordinal automata the labels are on one-step transitions and not on states as in standard Kripke structures usually used for stating LTL model checking. However, this is a superficial difference.

*2.4. A Non-elementary Decision Procedure*

Given LTL($\alpha$) models $\sigma, \sigma'$, we write $\sigma \approx_{\alpha'} \sigma'$ for some $\alpha' < \alpha$ whenever for every $\beta < \alpha'$, $\sigma(\beta) = \sigma'(\beta)$. Hence $\approx_\alpha$ is exactly the equality relation between LTL($\alpha$) models. Given a LTL($\alpha$) formula $\phi$, we write $\exp(\phi)$ to denote either $\alpha$ if $\alpha$ occurs in $\phi$ or the smallest ordinal of the form $\omega^\beta$ such that for every ordinal $\beta'$ occurring in $\phi$, $\beta' < \omega^\beta$.

**Lemma 2** *Let $\alpha$ be an ordinal closed under addition and $\phi$ be an LTL($\alpha$) formula. If $\sigma \in \mathrm{Mod}(\phi)$ and $\sigma \approx_{\exp(\phi)} \sigma'$, then $\sigma' \in \mathrm{Mod}(\phi)$.*

The proof of Lemma 2 is by an easy verification by observing that $\phi$ does not constraint states on positions greater than $\exp(\phi)$.

**Proposition 1** *Satisfiability for* LTL($\omega^\alpha$), $0 \leq \alpha \leq \omega$, *is decidable.*

The proof of Proposition 1 can be found in Appendix B and it provides a non-elementary complexity upper bound (a consequence of [35]). Furthermore, unlike the translation for LTL into the first-order theory of $\langle \omega, < \rangle$, the above translation makes a substantial use of second-order quantification. A translation into first-order logic has been found recently [14] but whether LTL($\omega^\omega$) has an elementary bound is still open. In the sequel, we considerably improve the bound for logics LTL($\omega^k$), $k \in \mathbb{N} \setminus \{0\}$, by using an automata-based approach.

**3. Automata-based Approach**

In this section, we show how to construct an ordinal automaton $\mathcal{A}_\phi$ such that its set of accepted words is precisely the models of $\phi$, extending the approach for LTL from [45]. In the rest of this section, $\phi \in$ LTL($\omega^k$) for some $k \geq 1$.

*3.1. Ordinal Automata*

We define ordinal automata as a generalization of Muller automata.

**Definition 1 (Ordinal Automaton)** *An ordinal automaton is a tuple $\langle Q, \Sigma, \delta, E, I, F \rangle$ where:*

- *$Q$ is a finite set of states, $\Sigma$ is a finite alphabet,*
- *$\delta \subseteq Q \times \Sigma \times Q$ is a one-step transition relation,*
- *$E \subseteq 2^Q \times Q$ is a limit transition relation,*
- *$I \subseteq Q$ is a finite set of initial states, $F \subseteq Q$ is a finite set of final states.*

We write $q \xrightarrow{a} q'$ whenever $\langle q, a, q' \rangle \in \delta$ and $q \to q'$ iff $q \xrightarrow{a} q'$ for some $a \in \Sigma$. A path of length $\alpha + 1$ is a map $r : \alpha + 1 \to Q$ such that

- for every $\beta \in \alpha$, $r(\beta) \to r(\beta + 1)$,
- for every limit ordinal $\beta \leq \alpha$, there is $P \to r(\beta) \in E$ s.t. $P = inf(\beta, r)$ with

$$inf(\beta, r) \stackrel{\text{def}}{=} \{q \in Q : \text{for every } \gamma \in \beta, \text{ there is } \gamma' \text{ such that } \gamma < \gamma' < \beta \text{ and } r(\gamma') = q\}.$$

A run of length $\alpha + 1$ is a path of length $\alpha + 1$ such that $r(0) \in I$. If $r(\alpha) \in F$ then $r$ is said to be accepting. The set of sequences recognized by the automaton $\mathcal{A}$, denoted by $L(\mathcal{A})$, is the set of $\alpha$-sequences $\sigma : \alpha \to \Sigma$ for which there is an accepting run $r$ of length $\alpha + 1$ verifying for every $\beta \in \alpha$, $r(\beta) \xrightarrow{\sigma(\beta)} r(\beta + 1)$.

Ordinal automata from Definition 1 are those defined in [28]. They are also exactly the $B'$-automata from [7, page 35], a variant of Wojciechowski's automata with no letter on limit transitions. The equivalence between these two formalisms is shown in [7, Sect. 2.5]. In [15, Def. 17], a similar notion is introduced and it is generalized in [11] to automata recognizing sequences over scattered linear orderings.

*Example.* We present below an example of ordinal automaton $\mathcal{A}$ with limit transitions $\{0\} \to 1$ and $\{0, 1\} \to 2$.

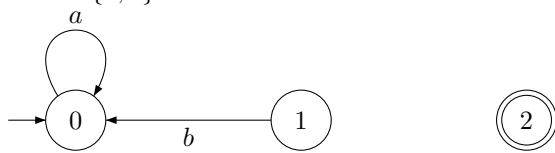

It is not difficult to show that $L(\mathcal{A})$ contains only $\omega^2$-sequences and $L(\mathcal{A}) = (a^\omega \cdot b)^\omega$.

*3.2. Synchronous Product*

We define below the synchronous product of two ordinal automata w.r.t. a synchronization alphabet. The purpose of this definition is to state the control problem in Section 5. Given two ordinal automata $\mathcal{A}_i = \langle Q_i, \Sigma_i, \delta_i, E_i, I_i, F_i \rangle$, for $i = 1, 2$, their synchronous product with respect to the set $X$ of synchronization vectors in $\Sigma_1 \times \Sigma_2 \times \Sigma$ ($\Sigma$ is a third arbitrary alphabet) is defined as the automaton $\mathcal{A}_1 \times_X \mathcal{A}_2 = \langle Q, \Sigma, \delta, E, I, F \rangle$ where:

- $Q = Q_1 \times Q_2$.
- $\langle q_1, q_2 \rangle \xrightarrow{c} \langle q'_1, q'_2 \rangle \in \delta$ iff there is $\langle a, b, c \rangle \in X$ such that $q_1 \xrightarrow{a} q'_1 \in \delta_1$, and $q_2 \xrightarrow{b} q'_2 \in \delta_2$.
- $P \to \langle q_1, q_2 \rangle \in E$ iff there exist $P_1 \to q_1 \in E_1$ and $P_2 \to q_2 \in E_2$ such that $\{q : \langle q, q' \rangle \in P\} = P_1$ and $\{q' : \langle q, q' \rangle \in P\} = P_2$.
- $I = I_1 \times I_2$, $F = F_1 \times F_2$.

By default, we write $\mathcal{A}_1 \times \mathcal{A}_2$ instead of $\mathcal{A}_1 \times_{ID} \mathcal{A}_2$ for the synchronized product with $\Sigma_1 = \Sigma_2 = \Sigma$ and $ID = \{\langle a, a, a \rangle : a \in \Sigma\}$.

**Proposition 2** *Let $\mathcal{A}_1$ and $\mathcal{A}_2$ be ordinal automata over the alphabet $\Sigma = \Sigma_1 = \Sigma_2$. We have $\mathrm{L}(\mathcal{A}_1) \cap \mathrm{L}(\mathcal{A}_2) = \mathrm{L}(\mathcal{A}_1 \times \mathcal{A}_2)$.*

*3.3. Hintikka Sequences*

We define below a notion of closure which generalizes the Fischer-Ladner closure [23].

**Definition 2 (Closure)** *The closure of $\phi$, denoted by $cl(\phi)$, is the smallest set of $\mathrm{LTL}(\omega^k)$ formulae such that*

- $\bot, \phi \in cl(\phi)$,
- $\neg \psi \in cl(\phi)$ *implies* $\psi \in cl(\phi)$,
- $\psi \in cl(\phi)$ *implies* $\neg \psi \in cl(\phi)$ *(we identify $\neg \neg \psi$ with $\psi$)*,
- $\psi_1 \wedge \psi_2 \in cl(\phi)$ *implies* $\psi_1, \psi_2 \in cl(\phi)$,
- $\mathtt{X}^\beta \psi \in cl(\phi)$ *and* $\beta \geq \omega^n$ *($0 \leq n < k$) imply* $\mathtt{X}^{\beta - \omega^n} \psi \in cl(\phi)$,
- $\psi_1 \mathtt{U}^\beta \psi_2 \in cl(\phi)$ *and* $\beta \geq \omega^n$ *($0 \leq n \leq k$) imply the formulae below belong to $cl(\phi)$: $\psi_1$, $\psi_2$, $\mathtt{X}^{\omega^n}(\psi_1 \mathtt{U}^{\beta - \omega^n} \psi_2)$, $\top \mathtt{U}^{\omega^n} \neg \psi_1$, $\psi_1 \mathtt{U}^{\omega^n} \psi_2$.*

It is not difficult to show that the notion of closure introduced above generalizes what is done for LTL.

**Lemma 3** *Let $\phi$ be an $\mathrm{LTL}(\omega^k)$ formula for some $k \geq 1$.*

**(I)** *There exists a polynomial such that $\mathrm{card}(cl(\phi))$ is in $2^{\mathcal{O}(p(|\phi|))}$ [resp. $\mathrm{card}(cl(\phi))$ is in $\mathcal{O}(p(|\phi|))$] when integers are encoded in binary [resp. in unary].*

**(II)** *For all $\mathtt{X}^\beta \psi \in cl(\phi)$ and $\gamma \leq \beta$, $\mathtt{X}^{\beta - \gamma} \psi \in cl(\phi)$.*

**(III)** *For all $\psi_1 \mathtt{U}^\beta \psi_2 \in cl(\phi)$ and $\gamma \leq \beta$, $\psi_1 \mathtt{U}^{\beta - \gamma} \psi_2 \in cl(\phi)$.*

From a formula $\phi$, we build an ordinal automata $\mathcal{A}_\phi$ such that $\mathrm{L}(\mathcal{A}_\phi)$ is precisely the set of $\mathrm{LTL}(\omega^k)$ models satisfying $\phi$. Following [45], the states of $\mathcal{A}_\phi$ are subsets of $cl(\phi)$ containing formulae to be satisfied in the future, including the current position. Hence, $cl(\phi)$ is built in such a way that if either $q' \to q$ or $P \to q$ are transitions in $\mathcal{A}_\phi$, then all the formulae to be satisfied in $q$ depending on $q'$ and $P$ are part of $cl(\phi)$.

**Definition 3** *A set $X \subseteq cl(\phi)$ is said to be locally maximally consistent with respect to $\phi$ iff it satisfies the conditions below:*

**(mc1)** $\bot \notin X$,

**(mc2)** *for every $\psi \in cl(\phi)$, $\psi \in X$ iff $\neg \psi \notin X$,*

**(mc3)** *for every $\psi_1 \wedge \psi_2 \in cl(\phi)$, $\psi_1 \wedge \psi_2 \in X$ iff $\psi_1, \psi_2 \in X$,*

**(mc4)** *for every $\mathtt{X}^0 \psi \in cl(\phi)$, $\mathtt{X}^0 \psi \in X$ iff $\psi \in X$,*

**(mc5)** *for every $\psi_1 \mathtt{U}^0 \psi_2 \in cl(\phi)$, $\psi_1 \mathtt{U}^0 \psi_2 \notin X$,*

**(mc6)** *for all $\psi_1 \mathtt{U}^\beta \psi_2 \in cl(\phi)$ and $\beta \geq \omega^n \geq 1$, $\psi_1 \mathtt{U}^\beta \psi_2 \in X$ iff either $\psi_1 \mathtt{U}^{\omega^n} \psi_2 \in X$ or $\neg(\top \mathtt{U}^{\omega^n} \neg \psi_1), \mathtt{X}^{\omega^n}(\psi_1 \mathtt{U}^{\beta - \omega^n} \psi_2) \in X$,*

**(mc7)** *for all $\psi_1 \mathtt{U}^\beta \psi_2, \psi_1 \mathtt{U}^{\beta'} \psi_2 \in cl(\phi)$ with $\beta \leq \beta'$, $\psi_1 \mathtt{U}^\beta \psi_2 \in X$ implies $\psi_1 \mathtt{U}^{\beta'} \psi_2 \in X$,*

**(mc8)** *for every $\psi_1 \mathtt{U}^1 \psi_2 \in cl(\phi)$, $\psi_1 \mathtt{U}^1 \psi_2 \in X$ iff $\psi_2 \in X$.*

Although all these conditions are used in the forthcoming proofs, at the moment we ignore whether Condition (mc7) is a consequence of the other conditions. We denote by $maxcons(\phi)$ the set of locally maximally consistent subsets of $cl(\phi)$.

For standard LTL, an Hintikka sequence $\rho$ for a formula $\phi$ is an $\omega$-sequence of sets of subformulae of $\phi$ such that $\phi$ is satisfiable iff $\phi$ has an Hintikka sequence. Local conditions in $\rho$ between two successive elements of the sequence are easy to handle in Büchi automata with the transition relation. The only global condition, stating that if $\psi_1 \mathtt{U} \psi_2$ occurs in the sequence, then some future element in the sequence contains $\psi_2$, is handled by the Büchi acceptance condition. Sometimes the non-uniform treatment between local conditions and the global condition is the source of confusion. The Hintikka sequences defined below are based on a similar principle except that we can extend advantageously the notion of locality. The Hintikka sequences $\rho$ are of the form $\rho : \omega^k \to 2^{cl(\phi)}$. Encoding conditions between $\rho(\beta)$ and $\rho(\beta + 1)$ can be performed by one-step transitions in ordinal automata. However, the presence of limit transitions allows us also to admit conditions between $\rho(\beta)$ and $\rho(\beta + \omega^{n'})$ with $0 \leq n' < k$. Hence, the global condition in Hintikka sequences of LTL formulae is replaced by a condition between $\rho(\beta)$ and $\rho(\beta + \omega)$. For transfinite sequences, the local and global conditions can be treated uniformly.

**Definition 4 (Hintikka Sequence)** *An Hintikka sequence for $\phi$ is a sequence $\rho : \omega^k \to 2^{cl(\phi)}$ such that*

**(hin1)** $\phi \in \rho(0)$,

**(hin2)** *for every $\beta < \omega^k$, $\rho(\beta) \in maxcons(\phi)$,*

**(hin3)** *for all $\beta < \omega^k$, $\mathtt{X}^{\beta'} \psi \in cl(\phi)$ and $0 \leq n' < k$ such that $\beta' \geq \omega^{n'}$, $\mathtt{X}^{\beta'} \psi \in \rho(\beta)$ iff $\mathtt{X}^{\beta' - \omega^{n'}} \psi \in \rho(\beta + \omega^{n'})$,*

**(hin4)** *for all $\beta < \omega^k$ and $\psi_1 \mathtt{U}^{\beta'} \psi_2 \in cl(\phi)$, (A) $\psi_1 \mathtt{U}^{\beta'} \psi_2 \in \rho(\beta)$ iff (B) there is $\beta \leq \beta'' < \beta + \beta'$ such that $\psi_2 \in \rho(\beta'')$ and for every $\beta \leq \gamma < \beta''$, $\psi_1 \in \rho(\gamma)$.*

Given a model $\sigma : \omega^k \to 2^{AP}$ and $\phi$ an LTL($\omega^k$) formula, we write $seq(\sigma, \phi)$ to denote the sequence $seq(\sigma, \phi) : \omega^k \to 2^{cl(\phi)}$ such that for every $\beta < \omega^k$, $seq(\sigma, \phi)(\beta) \stackrel{def}{=} \{\psi \in cl(\phi) : \sigma, \beta \models \psi\}$.

**Lemma 4** *Let $\sigma$ be a model such that $\sigma, 0 \models \phi$. Then $seq(\sigma, \phi)$ is an Hintikka sequence for $\phi$.*

The proof is by an easy verification.

**Lemma 5** *Let $\phi$ be a formula and $\rho : \omega^k \to 2^{cl(\phi)}$ be an Hintikka sequence. For every $\beta < \omega^k$, for every $\psi \in cl(\phi)$, $\psi \in \rho(\beta)$ iff $\sigma, \beta \models \psi$ where $\sigma : \omega^k \to 2^{AP}$ with $\sigma(\beta) = AP \cap \rho(\beta)$.*

The proof of Lemma 5 can be found in Appendix C. As a consequence of Lemma 4 and Lemma 5, we obtain the following proposition.

**Proposition 3** *$\phi$ is LTL($\omega^k$) satisfiable iff $\phi$ has an Hintikka sequence.*

*3.4. Automaton Construction*

We build an ordinal automaton $\mathcal{A}_\phi$ that recognizes only words of length $\omega^k$ over the alphabet $2^{AP}$ (assuming that AP is the finite set of atomic propositions occurring in $\phi$).

As the automata built from LTL formulae, states of $\mathcal{A}_\phi$ are locally maximally consistent sets. Each formula in a state has to be satisfied at the current position and this induces requirements for the future states of the run. Typically, if $\mathtt{X}^1 \psi$ belongs to some state, then the next state obtained by a one-step transition should contain the subformula $\psi$. However, the states in $\mathcal{A}_\phi$ are also made of some $n \in \{0, \ldots, k\}$ in order to remember the tail of the position of the state in the run. This stratification of states is useful for defining limit transitions and this is possible only because $k$ is strictly less than $\omega$.

The automaton $\mathcal{A}_\phi = \langle Q, \Sigma, \delta, E, I, F \rangle$ is defined as follows:

- $\Sigma = 2^{AP}$, $Q = maxcons(\phi) \times \{0, \ldots, k\}$,
- $I = \{\langle X, 0 \rangle \in Q : \phi \in X\}$, $F = \{\langle X, n \rangle \in Q : n = k\}$,
- $\langle X, n \rangle \xrightarrow{a} \langle X', n' \rangle \in \delta$ iff (one-step transition)

  **(A1)** $n < k$ and $n' = 0$,
  **(A2)** $X \cap AP = a$,
  **(A3)** for every $\mathtt{X}^\beta \psi \in cl(\phi)$ such that $\beta \geq 1$, $\mathtt{X}^\beta \psi \in X$ iff $\mathtt{X}^{\beta-1} \psi \in X'$.

- In order to define $E$, we introduce preliminary definitions. For every $\psi_1 \mathtt{U}^\beta \psi_2 \in cl(\phi)$, we write $P_{\psi_1 \mathtt{U}^\beta \psi_2}$ to denote the set below:

  $$\{\langle X, n \rangle : \text{ either } \psi_2 \in X \text{ or } \neg(\psi_1 \mathtt{U}^\beta \psi_2) \in X\}.$$

  For every $\langle X, n \rangle \in Q$ we write $Q_{\langle X, n \rangle}$ to denote the subset of $Q$ such that for every $\langle X', n' \rangle \in Q$, $\langle X', n' \rangle \in Q_{\langle X, n \rangle} \stackrel{def}{\Leftrightarrow}$

  **(A4)** $n' < n$,

**(A5)** for every $\mathtt{X}^\alpha \psi \in cl(\phi)$ with $\alpha \geq \omega^n$, $\mathtt{X}^\alpha \psi \in X'$ iff $\mathtt{X}^{\alpha - \omega^n} \psi \in X$.

For every $\langle X, n \rangle \in Q$, $Z \to \langle X, n \rangle \in E$ iff

**(A6)** $n \geq 1$,
**(A7)** $Z \subseteq Q_{\langle X, n \rangle}$,
**(A8)** $Z$ contains a state of the form $\langle Y, n-1 \rangle$,
**(A9)** for all $\psi_1 \mathtt{U}^\beta \psi_2 \in cl(\phi)$ and $\beta \geq \omega^n$ such that $\neg(\psi_1 \mathtt{U}^{\beta - \omega^n} \psi_2) \in X$, $P_{\psi_1 \mathtt{U}^\beta \psi_2} \cap Z \neq \emptyset$.

For LTL($\omega$), the above construction roughly corresponds to the Muller automaton obtained from the generalized Büchi automaton for the LTL formula $\phi$.

A state $\langle X, n \rangle \in Q$ is said to be of level $n$. Because of the strict discipline on levels in $\mathcal{A}_\phi$ it is not difficult to show the following result.

**Lemma 6** *Let $r : \omega^k + 1 \to Q$ be a run of $\mathcal{A}_\phi$. For every $\alpha < \omega^k + 1$, $r(\alpha)$ is of level $tail(\alpha)$.*

It remains to prove the main lemma whose proof requires some careful analysis. Indeed, it is the place where the conditions of the form (mc⋆) and (A⋆) are technically justified.

**Lemma 7** $\mathrm{Mod}(\phi)$ *is non-empty iff* $\mathrm{L}(\mathcal{A}_\phi)$ *is non-empty.*

The proof of Lemma 7 can be found in Appendix D. The automaton $\mathcal{A}_\phi$ has $2^{2^{\mathcal{O}(|\phi|)}}$ states and $2^{2^{2^{\mathcal{O}(|\phi|)}}}$ transitions. By [16, Proposition 6], the emptiness problem for ordinal automata is in P. So checking whether $\mathcal{A}_\phi$ accepts at least one word can be done in triple exponential time, which provides an elementary bound but not optimal as shown in Section 4.

**Proposition 4** $\mathrm{L}(\mathcal{A}_\phi) = \mathrm{Mod}(\phi)$.

Even though LTL($\omega^\omega$) is decidable (by translation into the monadic second-order theory of $\omega^\omega$), the proof of Lemma 7 cannot be extended to LTL($\omega^\omega$). Indeed, by [46, Sect. 8] (see also [28, Theorem 5.6]), there is no ordinal automaton accepting the language $\{a^\alpha\}$ for any countable ordinal $\alpha$ greater than or equal to $\omega^\omega$. However, for LTL($\omega^\omega$) it is open whether there exists a systematic construction of automata from formulae that allows to state a result as Lemma 7 (only equivalence of non-emptiness is required).

## 4. Computational Complexity

In this section, we show complexity results about satisfiability of LTL($\omega^k$).

*4.1. EXPSPACE-hardness*

Lemma 8 below states that although LTL and LTL($\omega$) are expressively equivalent, LTL($\omega$) is more concise than LTL mainly because $\mathtt{X}^n p$ is exponentially more succinct than $\overbrace{\mathtt{X} \cdots \mathtt{X}}^{n \text{ times}} p$ when $n$ is encoded in binary.

**Lemma 8** *Satisfiability for* LTL($\omega$) *is* EXPSPACE-*complete.*

Lemma 8 seems to contradict that LTL satisfiability is only PSPACE-complete but $X^{2^n} p$ can be represented with only $\mathcal{O}(n)$ bits. We prove the EXPSPACE-hardness since it will be used also for characterizing the complexity of LTL($\omega^\omega$). The proof is an adaptation of the proof of [27, Theorem 4.7] showing the PSPACE-hardness of LTL by reducing a PSPACE-complete tiling problem. In the case the natural numbers are encoded in unary in LTL($\omega$), we regain the PSPACE-completeness (see e.g. Section 3). The proof of Lemma 8 can be found in Appendix E.

As a consequence of Lemma 2 we obtain the following lower bound.

**Theorem 1** *For every ordinal $\alpha \geq 1$, satisfiability for LTL($\omega^\alpha$) is EXPSPACE-hard.*

*4.2. Succinct Ordinal Automata of Level k*

In order to refine the complexity result from Sect. 3, we define below specialized ordinal automata that recognize $\omega^k$-sequences. Similar automata can be found in the literature, see e.g. [18, 28, 7]. The main merit of the definition below is to allow easy manipulation in the forthcoming proofs.

**Definition 5 (Ordinal Automaton of Level $k$)** *An ordinal automaton $\mathcal{A} = \langle Q, \Sigma, \delta, E, I, F \rangle$ is said to be of level $k \geq 1$ iff there is a map $l : Q \to \{0, \ldots, k\}$ such that*

- *for every $q \in F$, $l(q) = k$;*
- *$q \xrightarrow{a} q' \in \delta$ implies $l(q') = 0$ and $l(q) < k$;*
- *$P \to q \in E$ implies*

  1. *$l(q) \geq 1$,*
  2. *for every $q' \in P$, $l(q') < l(q)$,*
  3. *there is $q' \in P$ such that $l(q') = l(q) - 1$.*

Hence, there is a partition of $Q$ of size $k + 1$ such that if $P \to q \in E$, then $max\{l(q') : q' \in P\} + 1 = l(q)$. Below, an ordinal automaton of level $k$ is denoted by $\langle Q, \Sigma, \delta, E, I, F, l \rangle$ where $l$ is the level function. Each set of states having the same level corresponds to a layer in Choueka's automata [18]. The automaton built in Section 3 is of level $k$ when the input formula is in LTL($\omega^k$). However, $\mathcal{A}_\phi$ is of triple [resp. double] exponential size in $|\phi|$ when integer are encoded in binary [resp. unary] which is still too much to characterize accurately the complexity of LTL($\omega^k$) satisfiability. That is why, we introduce below a special class of ordinal automata which can represent succinctly an exponential amount of limit transitions as the generalized Büchi automata can be viewed as a succinct representation of Muller automata. Hence, we shall construct $\mathcal{A}'_\phi$ such that $L(\mathcal{A}'_\phi) = L(\mathcal{A}_\phi)$, and $\mathcal{A}'_\phi$ is "only" of double [resp. simple] exponential size in $|\phi|$ when integers are encoded in binary [resp. unary].

**Definition 6 ($p(\cdot)$-Succinct Ordinal Automaton of Level $k$)** *Given a polynomial $p(\cdot)$, a $p(\cdot)$-succinct ordinal automaton of level $k$ is a structure $\mathcal{A} = \langle Q, \Sigma, \delta, E, I, F, l \rangle$ defined as an ordinal automaton of level $k$ except that $E$ is a set of tuples of the form $\langle P_0, P_1, \ldots, P_n, q \rangle$ with $n \geq 0$, $q \in Q$ and $P_0, \ldots, P_n \subseteq Q$ such that*

- $\langle P_0, P_1, \ldots, P_n, q \rangle \in E$ implies
    1. $1 \leq l(q) \leq k$,
    2. each state in $P_0$ is of level $l(q) - 1$,
    3. each state in $P_1 \cup \cdots \cup P_n$ is of level less than $l(q) - 1$,
    4. $n \leq p(|Q|)$,
- for every state $q$ of level strictly more than 0, there is at most one tuple in $E$ of the form $\langle P_0, P_1, \ldots, P_n, q \rangle$.

Each tuple $\langle P_0, P_1, \ldots, P_n, q \rangle$ encodes succinctly the set of limit transitions

$$trans(\langle P_0, P_1, \ldots, P_n, q \rangle) \stackrel{\text{def}}{=}$$

$$\{P \to q : P \subseteq Q, \ \forall i \ P_i \cap P \neq \emptyset \text{ and } \forall q' \in P, \ l(q') < l(q)\}.$$

Below, given a $p(\cdot)$-succinct ordinal automaton $\mathcal{A}$ of level $k$, we write $\mathcal{A}^o$ to denote the ordinal automaton of level $k$ $\langle Q, \Sigma, \delta, E', I, F, l \rangle$ with $E' = \bigcup_{t \in E} trans(t)$. The language recognized by $\mathcal{A}$ is defined as the language recognized by $\mathcal{A}^o$. In that way, a $p(\cdot)$-succinct ordinal automaton of level $k$ is simply a succinct encoding of some ordinal automaton of level $k$. An important property of such automata rests on the fact that the size of $E$ is in $\mathcal{O}(|Q|^2 \times p(|Q|))$. By contrast, in an ordinary ordinal automaton of level $k$, the cardinality of the set of limit transitions can be in the worst case exponential in $|Q|$.

The automaton $\mathcal{A}_\phi$ from Section 3.4 can be viewed as a $p_0(\cdot)$-succinct ordinal automaton of level $k$ with $p_0(x) = x$. Indeed, let $\mathcal{A}'_\phi$ be the $p_0(\cdot)$-succinct ordinal automaton of level $k$ defined as $\mathcal{A}_\phi$ with $l(\langle X, n \rangle) = n$ and $\langle P_0, P_1, \ldots, P_m, \langle X, n \rangle \rangle \in E$ iff

- $n \geq 1$,
- $P_0 \cup \cdots \cup P_m \subseteq Q_{\langle X, n \rangle}$ (see the definition of $Q_{\langle X, n \rangle}$ in Section 3.4),
- $P_0 = Q_{\langle X, n \rangle} \cap \{\langle Y, n' \rangle \in Q : n' = n - 1\}$.
- Let us pose $Z = \{\psi_1 \mathtt{U}^\beta \psi_2 \in cl(\phi) : \neg(\psi_1 \mathtt{U}^{\beta - \omega^n} \psi_2) \in X, \ \beta \geq \omega^n\}$. We have $|Z| = m$ and for every $\psi \in Z$, $\{\langle X, n \rangle : \text{ either } \psi_2 \in X \text{ or } \neg(\psi_1 \mathtt{U}^\beta \psi_2) \in X\} \in \{P_1, \ldots, P_m\}$ with $\psi = \psi_1 \mathtt{U}^\beta \psi_2$.

It remains to check that $m \leq |Q|$ (because of $p_0(\cdot)$). It is sufficient to observe that $m < |cl(\phi)|$ and $|Q|$ is in $\mathcal{O}(2^{|cl(\phi)|})$.

It is not difficult to show the following lemma:

**Lemma 9**

**(I)** $L(\mathcal{A}_\phi) = L(\mathcal{A}'_\phi)$.

**(II)** *In the unary [resp. binary] case, $\mathcal{A}'_\phi$ is of size exponential [resp. doubly exponential] in $|\phi|$ and requires only polynomial [resp. exponential] space to be built.*

### 4.3. Key Properties to Test Non-emptiness

In this section, we establish a few properties about runs in ordinal automata of level $k$.

**Lemma 10** *Let $\mathcal{A}$ be an automata of level $k$ and $r$ be a run of length $\alpha + 1$ with normal form $\beta + \omega^i \times n$, $n \geq 1$. Then $l(r(\beta + \omega^i \times n)) = i$.*

The proof is by an easy verification (induction on $i$). Lemma 10 is a slight generalization of Lemma 6. As a consequence, we obtain the following lemma.

**Lemma 11** *Let $\mathcal{A}$ be an automaton of level $k$. Then, its accepting runs are of length $\omega^k$.*

Lemma 12 below is the key property to obtain the NLOGSPACE upper bound for the non-emptiness problem of ordinal automata of level $k$, even in their succinct versions. It generalizes substantially the property that entails that the graph accessibility problem and the non-emptiness problem for generalized Büchi automata can be solved in non-deterministic logarithmic space. A Büchi automaton $\langle Q, \Sigma, \delta, I, F \rangle$ accepts a non-empty language iff there exists a path $q_0 \xrightarrow{n} q_f \xrightarrow{n'} q_f$ such that $q_0 \in I$, $q_f \in F$, $n \leq |Q|$ and $1 \leq n' \leq |Q|$. As usual, two states are in the relation $\xrightarrow{i}$ if there is a path of length $i$ between them. Similarly, a Muller automaton $\langle Q, \Sigma, \delta, I, \mathcal{F} \rangle$ accepts a non-empty language iff there exists a path $q_0 \xrightarrow{n} q_1 \to q_2 \to \cdots \to q_{n'}$ such that $q_0 \in I$, $q_1 = q_{n'}$, $\{q_1, \ldots, q_{n'}\} \in \mathcal{F}$ and $n' \leq |Q|^2$. Lemma 12 allows to generalize what is known about automata recognizing $\omega$-sequences: L($\mathcal{A}$) is non-empty iff $\mathcal{A}$ has an accepting run composed of a prefix followed by a loop with bounded length.

**Lemma 12** *Let $\mathcal{A}$ be an automaton of level $k$ and $r$ be a run of length $\omega^{k'} + 1$ for some $1 \leq k' \leq k$. Then, there is a run $r'$ of length $\omega^{k'} + 1$ such that*

- $r'(0) = r(0)$ and $r'(\omega^{k'}) = r(\omega^{k'})$,
- *there are $K \leq |Q|$ and $K' \leq |Q|^2$ such that for every $\alpha \geq \omega^{k'-1} \times K$ such that the normal form of $\alpha$ is $\omega^{k'-1} \times n + \beta$, $r'(\alpha) = r'(\omega^{k'-1} \times (n + K') + \beta)$.*

The proof of Lemma 12 can be found in Appendix F.

A consequence of Lemma 12 is that an automaton $\mathcal{A} = \langle Q, \Sigma, \delta, E, I, F, l \rangle$ of level $k$ accepts a non-empty language iff there exists a run $r : \omega^{k-1} \times (K + K') + 1 \to Q$ for some $K \leq |Q|$ and $K' \leq |Q|^2$ such that $r(\omega^{k-1} \times K) = r(\omega^{k-1} \times (K + K'))$ and

$$\{r(\beta) : \omega^{k-1} \times K \leq \beta \leq \omega^{k-1} \times (K + K')\} \to q_f \; \in \; E$$

for some state $q_f$ of level $k$.

More precisely, by taking $k' = k$ in Lemma 12, the automaton $\mathcal{A}$ accepts a non-empty language iff there are $K \leq |Q|$ and $K' \leq |Q|^2$ and $q_0^0, q_{k-1}^0, \ldots, q_0^{K+K'}, q_{k-1}^{K+K'}$ (these are landmark states of a run) such that

**(level$_0$)** $q_0^1, \ldots, q_0^{K+K'}$ are of level 0 and $q_0^0 \in I$.

**(level$_{k-1}$)** $q_{k-1}^0, \ldots, q_{k-1}^{K+K'}$ are of level $k-1$ and $q_{k-1}^K = q_{k-1}^{K+K'}$.

(**subruns**) for every $0 \leq i \leq K + K'$, there is a path $r_i : \omega^{k-1} + 1 \to Q$ such that $r_i(0) = q_0^i$ and $r_i(\omega_i^{k-1}) = q_{k-1}^i$. Each $r_i$ is part of a run from $(\omega^{k-1} \times i) + 1$ to $\omega^{k-1} \times (i+1)$.

(**k − 1 → 0**) for every $0 \leq i < K + K'$, $q_{k-1}^i \to q_0^{i+1}$.

(**last-transition**) $\{r_i(\beta) : K \leq i \leq K + K', 0 \leq \beta \leq \omega^{k-1}\} \to q_f \in E$ for some state $q_f$ of level $k$.

Existence of the runs $r_0, \ldots, r_{K+K'}$ above leads to the existence of an *accepting* run of length $\omega^k + 1$ as described below:

$$\underbrace{\overbrace{q_0^0 \ldots q_{k-1}^0}^{r_0} \to \overbrace{q_0^1 \ldots q_{k-1}^1}^{r_1} \cdots \cdots \overbrace{q_0^K \ldots q_{k-1}^K}^{r_K}}_{\text{prefix of length } \omega^{k-1} \times K} \to$$

$$\underbrace{(q_0^{K+1} \ldots \ldots q_{k-1}^{K+K'-1} \to \overbrace{q_0^{K+K'} \ldots q_{k-1}^{K+K'}}^{r_{K+K'}})^\omega}_{\text{loop of length } \omega^{k-1} \times K'}$$

In Condition (**subruns**), the existence of $r_i$ can be expressed recursively in a similar fashion on which is based the forthcoming algorithm to test non-emptiness. Even though is it obvious to see how the algorithm can work recursively, we have decided to provide the pseudo-code of the algorithm to underline some of its delicate aspects (in particular to get the proper amount of used space).

It is worth observing that Lemma 12 also holds for $p(\cdot)$-succinct ordinal automata of level $k$ since they form a special subclass of ordinal automata of level $k$. The succinctness of the representation of the set of limit transitions plays no role in Lemma 12.

*4.4. An Optimal Algorithm to Test Non-emptiness*

As seen earlier, non-emptiness is equivalent to the existence of some landmark states $q_0^0, q_{k-1}^0, \ldots, q_0^{K+K'}, q_{k-1}^{K+K'}$ satisfying the five above-mentioned conditions. In order to test non-emptiness of the language recognized by an automaton of level $k$, we introduce a function $acc(q, q')$ (see Fig. 1) that returns $\top$ iff there is a path $r$ of length $\omega^{l(q')} + 1$ such that $r(0) = q$ and $r(\omega^{l(q')}) = q'$. We design the following non-deterministic algorithm:

Non-empty?($\mathcal{A}$)

Guess $q_0 \in I$ and $q_f \in F$;

$acc(q_0, q_f)$.

Non-determinism is also present in the definition of $acc(q_0, q_f)$. A few global variables are used.

- The variables $\text{InLoop}_1, \ldots, \text{InLoop}_k$ are Boolean. Each variable $\text{InLoop}_i$ is equal to **true** iff the algorithm is guessing the periodic part of a run of length $\omega^i$ (which itself can belong to the periodic part of $\omega^j$ for some $j > i$). In particular, $\text{InLoop}_k$ is equal to true if the algorithm is building the periodic part of the global run, i.e. in the part $r_{K+1}, \ldots, r_{K+K'}$ according to the notations of Sect. 4.3. In that case, every state in the run from $q$ to $q'$ has to be recorded in order to be also able to fire the last limit transition (see Condition **(last-transition)** in Sect. 4.3).

- Moreover, for every $i \in \{1, \ldots, k\}$, the variable $\uparrow_i$ contains the address of the occurrence of a state in the left part of a rule $P \to q''$ with $l(q'') = i$: $\mathcal{O}(k \times log|\mathcal{A}|)$ bits are needed in total.

Remember that $\mathcal{A}$ is encoded as a string and the address of the occurrence of a state is simply a position in that string, which requires only $\mathcal{O}(log|\mathcal{A}|)$ bits. The variable $\uparrow_i$ is updated when the state whose address is $\uparrow_i$ is detected in the periodic part of the run.

In the definition of $acc(q, q')$, in order to test whether there is a path $r$ of length $\omega^{l(q')} + 1$ such that $l(q') \geq 1$, $r(0) = q$ and $r(\omega^{l(q')}) = q'$, Lemma 12 guarantees that the periodic part of $r$ is of length at most $\omega^{l(q')-1} \times |Q|^2$ and the prefix is of length at most $\omega^{l(q')-1} \times |Q|$. This explains the two main loops of $acc(q, q')$. The two "for" loops guess respectively the prefix and the period. Observe that the iteration variable $i$ is only used to guarantee that the lengths of the subruns are correct. When a state $t$ is guessed in the periodic part of the global run, one has to check that $t$ indeed belongs to rules of the form $P \to q''$ with $l(q'') > l(q_t)$ and one updates the variables $\uparrow_i$ since $t$ has been detected (see Fig. 2).

**Lemma 13** *Non-empty?($\mathcal{A}$) = $\top$ iff $L(\mathcal{A}) \neq \emptyset$.*

Let us briefly analyze the complexity of the algorithm. Global variables require $\mathcal{O}(k \times log|\mathcal{A}|)$ space and the recursive depth is at most $k$. By passing the variables by reference, the whole algorithm requires space $\mathcal{O}(k \times log|\mathcal{A}|)$.

As a consequence we obtain the following theorem.

**Theorem 2** *For every $k \geq 0$, the non-emptiness problem for ordinal automata of level $k$ is* NLOGSPACE*-complete.*

It is worth observing that as a corollary of [16], the non-emptiness problem for ordinal automata is in P. Herein, we refine this result for a subclass of ordinal automata: for every $k \geq 1$, the non-emptiness problem for ordinal automata of level $k$ is in NLOGSPACE. However, our algorithm runs in time $\mathcal{O}(|\mathcal{A}|^{2 \times k})$: in order to save space, we do not keep in memory the outcomes of previous accessibility checks (similarly to the proof establishing that logarithmic reductions are closed under composition). It is open whether the non-emptiness problem for ordinal automata of level $k$ for some $k \geq 0$ is P-hard ($k$ is not fixed).

**Corollary 1** *The non-emptiness problem for Muller automata ($k = 1$) is* NLOGSPACE*-complete.*

The NLOGSPACE upper bound is a consequence of Theorem 2 and the NLOGSPACE lower bound can be obtained by reducing the graph accessibility problem.

**Corollary 2** *For all $k \geq 0$ and polynomial $p(\cdot)$, the non-emptiness problem for $p(\cdot)$-succinct ordinal automata of level $k$ is* NLOGSPACE*-complete.*

$acc(q, q')$ $(l(q') \leq k, l(q) = 0)$

$k' := l(q') - 1$;

If $k' \geq 0$ then

**(initial-guesses)**
- $\text{InLoop}_{k'+1} := \texttt{false}$;
- Guess a rule $P \to q'$;
- $\uparrow_{k'+1}$ takes the value of the address of the first state in $P$;
- Guess $K \leq |Q|$ and $K' \leq |Q|^2$;

($\star$) $q_0 := q$;

**(guess-prefix)** For $i = 1$ to $K$ do
- Guess $q_{k'} \in P$ of level $k'$;
- Check&Update($q_{k'}$);
- If $acc(q_0, q_{k'})$ then guess $q_0$ such that $l(q_0) = 0$ and $q_{k'} \to q_0$ otherwise `abort`;

($\star$) If $q_{k'}^{\text{repeat}} := q_{k'}$ (forthcoming repeating state);

($\star$) $\text{InLoop}_{k'+1} = \texttt{true}$;

($\star$) Guess $q_{k'} \in P$ of level $k'$;

($\star$) Check&Update($q_0$);Check&Update($q_{k'}$);

**(guess-period)** For $i = 1$ to $K'$ do

If $acc(q_0, q_{k'})$ then
- Guess $q_0$ such that $l(q_0) = 0$ and $q_{k'} \to q_0$;
- $q_{k'}^{\text{aux}} := q_{k'}$;
- Guess $q_{k'} \in P$ of level $k'$;
- If $i \neq K'$ (not the last dummy guess) then (Check&Update($q_0$);Check&Update($q_{k'}$));

otherwise `abort`;

**(final-check)** If one of the conditions below fails then `abort` otherwise `accept`

**(C1)** $\uparrow_{k'+1} \neq nil$ (some state in $P$ has not been visited),
**(C2)** $q_{k'}^{\text{aux}} \neq q_{k'}^{\text{repeat}}$ (wrong choice of the repeating state of level $k'$)

otherwise if $q \to q'$ then `accept` otherwise `abort`.

Figure 1: Accessibility function

---

Check&Update($q$)

For $1 \leq i \leq k$ do

($q$ **is desirable**) If $\text{InLoop}_i = \texttt{true}$ and $\uparrow_i$ contains the address of an occurrence of $q$ in the left part of a rule then $\uparrow_i$ takes the value of the next state in the rule or *nil* if there is no such a remaining state;

($q$ **is undesirable**) If $\text{InLoop}_i = \texttt{true}$ and $l(q) \leq i - 1$ and $q$ does not occur in the left part of the rule that is currently pointed by $\uparrow_i$ then `abort`. (one needs another variable to visit the states in the left part of that rule)

`accept`.

---

Figure 2: Update of the variables $\uparrow_i$s

Indeed, for ordinary ordinal automata, rules $P \to q'$ in $E$ are guessed (see Fig. 1) whereas for $p(\cdot)$-succinct ordinal automata of level $k$, we guess which element for each $P_i$ occurring in $P_0, \ldots, P_m \to q'$ is repeated infinitely often. So we guess $q_0, \ldots, q_m \to q'$ and $\downarrow_{l(q')-1}$ contains the address of the occurrence of some $q_i$. Of course we do not guess $q_0, \ldots, q_m \to q'$ at once for space saving but rather guess each $q_i$ step by step. Because in succinct ordinal automata, we only specify the existence of states repeated infinitely often (as in generalized Büchi automata), the second condition can be deleted in Fig. 2.

*4.5. Optimal Complexity Upper Bounds*

We are now in position to characterize the computational complexity of satisfiability and model-checking problems.

**Theorem 3** *For every $k \geq 1$, the satisfiability problem for $\text{LTL}(\omega^k)$ is* PSPACE-*complete when the integers are encoded in unary and the problem is in* EXPSPACE-*complete when the integers are encoded in binary.*

The proof of Theorem 3 can be found in Appendix G. Another way to prove Theorem 3 suggested to us in [37] consists in showing that LTL with strict Since and Until over $\omega^\omega$-sequences is in PSPACE. Indeed, it is then possible to define concisely a formula $\varphi_i$ stating that the current position is a multiple of $\omega^i$ for $i \in \omega$. Our operators $\mathtt{U}^{\omega^i}$ and $\mathtt{X}^{\omega^i}$ for some $i \geq 1$ are then definable as follows: $\psi \mathtt{U}^{\omega^i} \psi' \sim \psi' \vee ((\neg \varphi_i \wedge \psi)\mathtt{U}(\neg \varphi_i \wedge \psi'))$ and $\mathtt{X}^{\omega^i} \psi \sim ((\neg \varphi_i)\mathtt{U}(\varphi_i \wedge \psi))$. Renaming of subformulae are necessary to guarantee that the translation can be performed in logarithmic space. It is however open whether for every countable ordinal $\alpha$, LTL with strict Since and Until over $\alpha$-sequences is in PSPACE.

Complexity of the model-checking problem for $\text{LTL}(\omega^k)$ can be now fully characterized.

**Theorem 4** *For every $k \geq 1$, the model-checking problem for $\text{LTL}(\omega^k)$ is* PSPACE-

*complete when the integers are encoded in unary and the problem is in* EXPSPACE-*complete when the integers are encoded in binary.*

The proof of Theorem 4 can be found in Appendix H and it is a slight variant of what exists for LTL. Theorem 4 can be refined by admitting succinct ordinal automata as inputs of the model-checking problem.

**Theorem 5** *For every $k \geq 1$, the model-checking problem for $\text{LTL}(\omega^k)$ restricted to x-succinct ordinal automata of level $k$ is* PSPACE-*complete when the integers are encoded in unary and the problem is* EXPSPACE-*complete when the integers are encoded in binary.*

Hence, even if the system is defined succinctly, the worst-case complexity remains identical. The proof of Theorem 5 can be found in Appendix I.

## 5. Application: Control of Physical Systems

In this section, we formalize the control problem of a physical system by a computer system by using ordinal automata and the logics $\text{LTL}(\omega^k)$. Even though it is the original motivation of our investigations on the logics $\text{LTL}(\alpha)$, at this point of the paper we have all the necessary definitions and results to state concisely the problem. Physical systems are often modelled by differential equations. Solving those equations can then involve computations of limits. For example, the law of movement of a bouncing ball implies that, when it is lifted-up, it will bounce an infinite number of times in a finite amount of time. It can be seen as a Zeno sequence of actions. We model a system by an ordinal automaton recognizing $\omega^k$-sequences. For instance, the law of movement of the bouncing ball corresponds to $\omega^2$-sequences and the set of acceptable behaviors of the ball is modelled by a set of sequences of the same length $\omega^2$. On the other hand, the controller is an operational model working on $\omega$-sequences.

Before stating the control problem, we need to give definitions about the way to transform an ordinal automaton of level 1 into an ordinal automaton of level $k \geq 2$ that has relevant actions only on states in positions of the form $\omega^{k-1} \times n$ (*lifting*). As usual, $\text{LTL}(\omega^k)$ formulae can be viewed equivalently as ordinal automata of level $k$ and we shall use these different representations depending on the context (see [3] for a similar standard treatment between formulae and automata).

*5.1. Lifting*

In order to synchronize the system $\mathcal{S}$ with a controller working on $\omega$-sequences, we need to transform the controller so that its product with $\mathcal{S}$ only constraints states on positions $\omega^{k-1} \times n$, $n \in \mathbb{N}$. The other positions are not constrained.

**Definition 7 (Lifting)** *Let $\mathcal{A} = \langle Q, \Sigma, \delta, E, I, F, l \rangle$ be an automaton of level 1 (the final states are the only states of level 1). We define its lifting $\text{lift}_k(\mathcal{A})$ at level $k \geq 2$ to be the automaton $\langle Q', \Sigma, \delta', E', I', F, l' \rangle$ by:*

- $Q' = (\{0, \ldots, k-1\} \times (Q \setminus F)) \cup F$, $I' = \{k-1\} \times I$,
- $l'(q) = k$ *for* $q \in F$ *and* $l'(\langle i, q' \rangle) = i$,

- $\delta' = \{\langle k-1, q\rangle \xrightarrow{a} \langle 0, q'\rangle \,:\, q \xrightarrow{a} q' \in \delta\} \cup$
  $\{\langle i, q\rangle \xrightarrow{a} \langle 0, q\rangle \,:\, 0 \leq i < k,\ a \in \Sigma,\ q \notin F\}$ ,
- $E' = \{\{\langle 0, q\rangle, \ldots, \langle i-1, q\rangle\} \to \langle i, q\rangle \,:\, 1 \leq i < k,\ q \in Q\} \cup \{\{\langle 0, q_1\rangle, \ldots, \langle k-1, q_1\rangle, \ldots, \langle 0, q_n\rangle, \ldots, \langle k-1, q_n\rangle\} \to q \mid \{q_1, \ldots q_n\} \to q \in E\}$.

*Example.* We present below an example of ordinal automaton $\mathcal{A}$ with limit transition $\{q_0, q_1, q_2\} \to q_3$ and the corresponding automaton $lift_2(\mathcal{A})$ with limit transitions $\{\langle 0, q_0\rangle\} \to \langle 1, q_0\rangle$, $\{\langle 0, q_1\rangle\} \to \langle 1, q_1\rangle$, $\{\langle 0, q_2\rangle\} \to \langle 1, q_2\rangle$, and
$\{\langle 0, q_0\rangle, \langle 1, q_0\rangle, \langle 0, q_1\rangle, \langle 1, q_1\rangle, \langle 0, q_2\rangle, \langle 1, q_2\rangle\} \to q_3$.

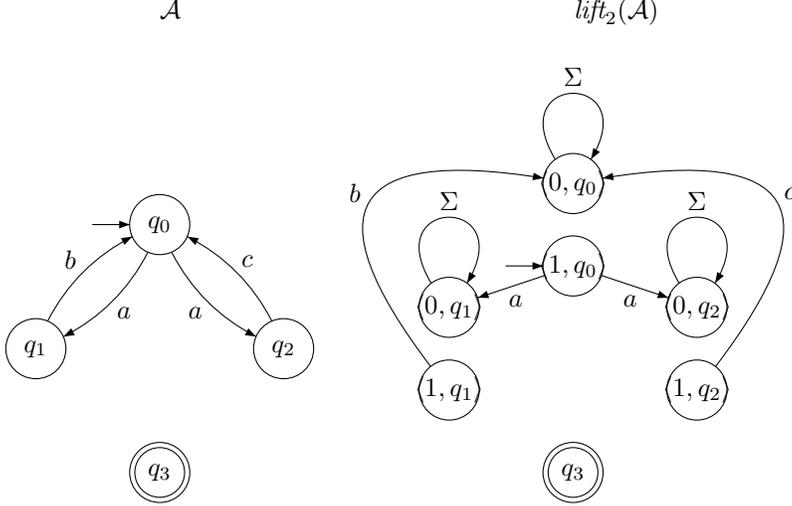

$\mathcal{A}$ $\qquad\qquad lift_2(\mathcal{A})$

**Proposition 5** *For all $w \in \Sigma^{\omega^k}$, $w \in \mathrm{L}(lift_k(\mathcal{A}))$ iff the word $w' \in \Sigma^\omega$, defined by $w'(i) = w(\omega^{k-1} \times i)$, is in $\mathrm{L}(\mathcal{A})$.*

The proof of Proposition 5 can be found in Appendix J.

*5.2. The Control Problem*

**Definition 8 (Physical system)** *A physical system $\mathcal{S}$ is modelled as a structure $\langle \mathcal{A}, Act_C, Act_O, Act\rangle$ where*

- $\mathcal{A}$ *is an ordinal automaton of level $k$ with alphabet $2^{Act}$ where $Act$ is a finite non-empty set of actions,*
- $Act_O \subseteq Act$ *is the set of observable actions,*
- $Act_C \subseteq Act_O$ *is the set of controllable actions. The set of uncontrollable actions is denoted by $Act_{nc}$.*

A specification of the system $\mathcal{S}$ is naturally an $\mathrm{LTL}(\omega^k)$ formula $\psi$. A controller $\mathcal{C}$ for the pair $\langle \mathcal{S}, \psi\rangle$ is a system whose complete executions are $\omega$-sequences (typically ordinal automata of level 1) verifying the properties below.

- Only observable actions are present in the controller. Hence, thanks to the synchronization mode, in the product system between $\mathcal{S}$ and $\mathcal{C}$, unobservable actions do not change the $\mathcal{C}$-component of the current state. So,

- **(obs)** the alphabet of $\mathcal{C}$ is $2^{Act_O}$ and for every state $q$ of $\mathcal{C}$, there is a transition $q \xrightarrow{\emptyset} q$.

- From any state of $\mathcal{C}$, uncontrollable actions can always be executed:

  **(unc)** $\forall q \cdot \forall a \subseteq Act_O \setminus Act_C$, there is a transition $q \xrightarrow{b} q'$ in $\mathcal{C}$ such that $b \cap Act_{nc} = a$.

- Finally, the system $\mathcal{S}$ controlled by $\mathcal{C}$ satisfies $\psi$. Because $\mathcal{S}$ and $\mathcal{C}$ work on sequences of different length, the controlled system is in fact equal to $lift_k(\mathcal{C}) \times_Y \mathcal{S}$ for some set $Y$ of synchronization vectors. $\mathcal{S}$ and $\mathcal{C}$ synchronize on observable actions:

  **(syn)** $Y = \{\langle X, X', X'' \rangle \in Act \times Act_O \times Act : X \cap Act_O = X', \ X = X''\}$.

  This is equivalent to check the emptiness of the language of the product automaton $(\mathcal{S} \times_Y lift_k(\mathcal{C})) \times \mathcal{A}_{\neg\psi}$.

Hence, the *control problem for* $\text{LTL}(\omega^k)$ is defined as follows:

**input:** a system $\mathcal{S} = \langle \mathcal{A}, Act_C, Act_O, Act \rangle$ with ordinal automaton $\mathcal{A}$ of level $k$ and an $\text{LTL}(\omega^k)$ formula $\phi$ over atomic formulae in $Act$.

**output:** is there an ordinal automaton $\mathcal{C}$ of level 1 satisfying (obs) and (unc) and such that all the words of length $\omega^k$ accepted by $\mathcal{S} \times_Y lift_k(\mathcal{C})$ satisfy $\phi$ with $Y$ verifying (syn).

It is worth noting that the lifting construction oversimplifies the physical synchronization between the system and the controller. Indeed, the fact that $lift_k(\mathcal{C})$ synchronizes with $\mathcal{S}$ every $\omega^{k-1}$ step idealizes the ability of the controller. Assuming that $\mathcal{C}$ interacts with $\mathcal{S}$ at the steps $0 < \alpha_1 < \alpha_2 < \ldots$ with $lim_{i \to w} \alpha_i = \omega^k$ is more realistic. With the construction $lift_k(\mathcal{C})$, it is implicitly assumed that $\alpha_i$ is precisely $\omega^{k-1} \times i$.

The very complexity of the control problem is open (see related results in the recent [14]) but as a consequence of Theorem 4 we obtain the following result.

**Proposition 6** *The problem of checking whether the language accepted by* $(\mathcal{S} \times_Y lift_k(\mathcal{C})) \times \mathcal{A}_{\neg\psi}$ *is non-empty, given a physical system* $\mathcal{S}$, *a controller* $\mathcal{C}$ *and a specification* $\psi$ *is decidable.*

We explained how to check that a controller is correct with respect to a specification, but we do not address here the controller synthesis issue.

*5.3. Example*

Consider the system is a bouncing ball [20] with three actions *lift-up*, *bounce* and *stop*, where only *lift-up* is controllable, and only *stop* and *lift-up* are observable. The law of the ball is described by the following $\text{LTL}(\omega^2)$ formula:

$$\phi = \mathtt{G}^{\omega^2}(\textit{lift-up} \Rightarrow \mathtt{X}^1(\mathtt{G}^\omega \textit{bounce} \wedge \mathtt{X}^\omega \textit{stop}))$$

$\mathtt{G}^\alpha \varphi$ is an abbreviation for $\neg(\top \mathtt{U}^\alpha \neg\varphi)$. Informally, $\phi$ states that when the ball is lifted-up, then it bounces an infinite number of times in a finite time and then stops. An equivalent ordinal automaton $\mathcal{A}_\phi$ working on $\omega^2$-sequences can be easily defined. The specification is given by the following LTL($\omega^2$) formula:

$$\psi = \mathtt{G}^{\omega^2} \mathtt{X}^1 bounce$$

Informally, $\psi$ states that the ball should almost always be bouncing.

A possible controller for this system is described by the following LTL formula:

$$\varphi = \text{lift-up} \wedge \mathtt{G}^\omega(\text{stop} \Rightarrow \text{lift-up})$$

Informally, $\varphi$ states that the controller should lift-up the ball at the beginning and then lift-up it again each time it stops. Similarly, an equivalent ordinal automaton $\mathcal{A}_\varphi$ working on $\omega$-sequences can be easily defined.

## 6. Concluding Remarks

We have introduced a family of temporal logics to specify the behavior of systems by assuming that the sequence of actions is isomorphic to some well-ordered linear ordering (see the bouncing ball example in Section 5). Our aim is to control such physical systems by designing controllers that safely work on $\omega$-sequences but interact synchronously with the physical system in order to restrict their behaviors. We have extended linear-time temporal logic LTL to $\alpha$-sequences for any countable ordinal $\alpha$ closed under addition, by considering quantitative operators indexed by ordinals smaller than $\alpha$. This is a new class of linear-time temporal logics for which we have shown that LTL($\omega^\omega$) is decidable by reduction to the monadic second-order theory $\langle \omega^\omega, < \rangle$ and for every $k \geq 1$, LTL($\omega^k$) satisfiability problem is PSPACE-complete [resp. EXPSPACE-complete] when the integers are encoded in unary [resp. in binary] generalizing what is known about LTL. Our proof technique is inspired from [45] with significant extensions in order to deal with the interaction between arithmetics on ordinals and temporal operators. We have introduced a new class of succinct ordinal automata in order to fully characterize the complexity of the logics. The treatment of these aspects leads to the most difficult technical parts of the paper. Finally, the complexity results for satisfiability can be lifted to model-checking: the model checking problem for LTL($\omega^k$) is PSPACE-complete [resp. EXPSPACE-complete] when the integers are encoded in unary [resp. binary].

A lot of work remains to be done even though our logics working on $\omega^k$-sequences have been shown to admit reasoning tasks of complexity similar to that of LTL. Synthesis of controllers working on $\omega$-sequences on the line of Section 5 is on the top of our priority list as well as the search for well-motivated examples where ordinals greater than $\omega^2$ are needed. It is also natural to wonder whether LTL($\omega^\omega$) satisfiability is an elementary problem and whether for every countable ordinal $\alpha$, LTL($\alpha$) is decidable. Observe that the monadic second-order theory of every countable ordinal $\alpha$ is known to be decidable [13] but this theory has no addition and we need it in some way in LTL($\alpha$) to deal with the operators $\mathtt{X}^\beta$. Finally, LTL is known to be initially equivalent to the first-order theory of $\langle \omega, < \rangle$ by Kamp's theorem [33] and by the separation theorem [24]. Is LTL($\omega^k$) also initially equivalent to the first-order theory of $\langle \omega^k, < \rangle$? It is unlikely the case since by [31], the future fragment of MLO over the class of ordinals does not have the finite base property.

**Acknowledgments.** Thanks to Peter van Emde Boas and Eric Allender for pointing out us the existence of [6, Corollary 3.36] that happens to be a key argument in the proof of Theorem 3, to Paul Schupp and Paul Gastin for providing us essential parts of Rohde's thesis (we have now the whole document), to Claudine Picaronny for providing us articles on ordinal automata, to Nicolas Bedon for many electronical discussions, to Jean-Michel Couvreur for fruitful discussions. We would like to warmly thank Thierry Cachat for his comments on earlier versions of this work and for so helpful and fruitful discussions. We thank also Alexander Rabinovich for pointing us to [31] and for suggesting a simpler proof strategy for Theorem 3. Finally, we would like to thank the anonymous referees for their helpful remarks and suggestions on earlier versions.

## Appendix A: Proof of Lemma 1

First suppose that $\alpha = \omega^\beta$ and take $\beta_1 \leq \beta_2 < \omega^\beta$. The Cantor's normal form of $\beta_1$ [resp. $\beta_2$] is of the form $\omega^{\gamma_1} n_1 + \gamma_1'$ [resp. $\omega^{\gamma_2} n_2 + \gamma_2'$] with either $\gamma_2 > \gamma_1$ or $n_2 \geq n_1$. Hence,
$$\beta_1 + \beta_2 \leq \beta_2 + \beta_2 < \omega^{\gamma_2} \times n$$
for some $n > 1$. Consequently, $\beta_1 + \beta_2 < \omega^\beta$ since $\omega^{\gamma_2} \times n < \omega^{\gamma_2+1} \leq \omega^\beta$.

Now suppose that the Cantor normal form of $\alpha$ is $\omega^{\beta_1}.n_1 + \cdots + \omega^{\beta_p}.n_p$ where $p > 1$ and $n_p \neq 0$. The ordinals $\omega^{\beta_1}.n_1$ and $\omega^{\beta_1}$ are strictly less than $\alpha$, but their sum is strictly greater. $\square$

## Appendix B: Proof of Proposition 1

By Lemma 2, it is sufficient to show that $\mathrm{LTL}(\omega^\omega)$ is decidable. We extend the standard translation from LTL into the monadic second order theory of $\langle \omega, < \rangle$ in order to translate $\mathrm{LTL}(\omega^\omega)$ into the monadic second order theory of $\langle \omega^\omega, < \rangle$ since the monadic second order theory of $\langle \alpha, < \rangle$ for every countable ordinal $\alpha$ is decidable [13, Theorem 4.12]. The main difficulty rests on the definition of a formula $+_\beta(x, y)$ for some $\beta < \omega^\omega$ such that $\langle \omega^\omega, < \rangle \models_v +_\beta(x, y)$ with $v : \{x, y\} \to \omega^\omega$ iff $v(y) = v(x) + \beta$. The relation $\models_v$ is the standard satisfaction relation under the valuation $v$. It is worth observing that addition is not present in the monadic second order theory of $\langle \omega^\omega, < \rangle$. With the help of $+_\beta(x, y)$ we define a two-places map $t(\cdot, \cdot)$ such that for any $\mathrm{LTL}(\omega^\omega)$ formula $\phi$ built over the propositional variables $p_1, \ldots, p_n$, for any $\sigma : \omega^\omega \to 2^{\{p_1,\ldots,p_n\}}$, we have $\sigma, 0 \models \phi$ iff $\langle \omega^\omega, <, P_1, \ldots, P_n \rangle \models_v t(\phi, x_0)$ with $v(x_0) = 0$ and for $1 \leq l \leq n$, $P_l = \{\beta \in \omega^\omega : p_l \in \sigma(\beta)\}$.

- $t(p, x) = p(x)$, $t(\phi \wedge \psi, x) = t(\phi, x) \wedge t(\psi, x)$, $t(\neg \phi, x) = \neg t(\phi, x)$,
- $t(\mathtt{X}^\beta \phi, x) = \exists\, y\ +_\beta (x, y) \wedge t(\phi, y)$,
- $t(\phi \mathtt{U}^\beta \psi, x) = \exists\, y\, y'\ +_\beta (x, y') \wedge (x \leq y \wedge y < y') \wedge t(\psi, y) \wedge (\forall\, z\ (x \leq z \wedge z < y) \Rightarrow t(\phi, z))$ if $\beta < \omega^\omega$.
- $t(\phi \mathtt{U}^{\omega^\omega} \psi, x) = \exists\, y\ (x \leq y) \wedge t(\psi, y) \wedge (\forall\, z\ (x \leq z \wedge z < y) \Rightarrow t(\phi, z))$.

The formulae of the form $+_\beta(x, y)$ with $\beta < \omega^\omega$ are inductively defined as follows:

1. $+_0(x, y) \stackrel{\mathrm{def}}{=} (x = y)$,
2. $+_1(x, y) \stackrel{\mathrm{def}}{=} \forall\, z\ (z > x \Rightarrow y \leq z) \wedge (x < y)$,
3. $+_{\omega^k n + \beta}(x, y) \stackrel{\mathrm{def}}{=} \exists\, z\ +_{\omega^k}(x, z) \wedge +_{\omega^k(n-1)+\beta}(z, y)$ ($n \geq 1$, $k \geq 0$),
4. $+_{\omega^k}(x, y)$ ($k \geq 1$) is defined as $\exists\, X\ \phi_1 \wedge \cdots \wedge \phi_6$ where the $\phi_i$s are defined as follows.

   (a) $\phi_1 \stackrel{\mathrm{def}}{=} \forall\, z\ z \in X \Rightarrow x < z$,
   (b) $\phi_2 \stackrel{\mathrm{def}}{=} \forall\, z, z'\ (z \in X \wedge +_{\omega^{k-1}}(z, z')) \Rightarrow z' \in X$,
   (c) $\phi_3 \stackrel{\mathrm{def}}{=} \exists\, z\ z \in X \wedge +_{\omega^{k-1}}(x, z)$,
   (d) $\phi_4 \stackrel{\mathrm{def}}{=} \forall\, z\ z \in X \Rightarrow z < y$,

(e) $\phi_5 \stackrel{\text{def}}{=} \forall z \ (\forall z' \ (z' \in X \Rightarrow z' < z)) \Rightarrow y \leq z$,

(f) $\phi_6 \stackrel{\text{def}}{=} \forall X' \ (X' \subset X) \Rightarrow \phi_2^{\neg} \vee \phi_3^{\neg}$ where the $\phi_i^{\neg}$'s are defined as follows.

   i. $\phi_2^{\neg} \stackrel{\text{def}}{=} \neg(\forall z, z' \ (z \in X' \wedge +_{\omega^{k-1}}(z, z')) \Rightarrow z' \in X')$,
   
   ii. $\phi_3^{\neg} \stackrel{\text{def}}{=} \neg(\exists z \ z \in X' \wedge +_{\omega^{k-1}}(x, z))$.

It is not difficult to show that the definition of $+_\beta(x,y)$ with $\beta < \omega^\omega$ is correct since the recursive steps involve only ordinals strictly less than $\beta$. Some explanations are in order. In (4.), the variable $X$ is enforced to be interpreted as the set $\{\gamma + \omega^{k-1}, \gamma + \omega^{k-1} \times 2, \gamma + \omega^{k-1} \times 3, \ldots\}$ where the variable $x$ is interpreted by $\gamma$. The value of $y$ is then the limit of this set. By satisfaction of $\phi_4$ and $\phi_5$, $y$ is interpreted as the least upper bound of $\{\gamma + \omega^{k-1}, \gamma + \omega^{k-1} \times 2, \gamma + \omega^{k-1} \times 3, \ldots\}$ which is precisely $\gamma + \omega^k$. The formula $\phi_6$ states that $X$ is interpreted as the smallest set satisfying the formula $\phi_1 \wedge \phi_2 \wedge \phi_3$. $\square$

## Appendix C: Proof of Lemma 5

Lemma C.1 below states a useful property about Hintikka sequences.

**Lemma C.1** Let $\rho : \omega^k \to 2^{cl(\phi)}$ be a Hintikka sequence for $\phi$. For all $\beta < \omega^k$ and $\mathtt{X}^{\beta'}\psi \in cl(\phi)$, $\mathtt{X}^{\beta'}\psi \in \rho(\beta)$ iff $\psi \in \rho(\beta + \beta')$.

**Proof.** By using (hin3), it is easy to show by induction that for all $\beta'' \leq \beta'$, $\mathtt{X}^{\beta'}\psi \in \rho(\beta)$ iff $\mathtt{X}^{\beta'-\beta''}\psi \in \rho(\beta + \beta'')$. Hence, $\psi \in \rho(\beta + \beta')$. $\square$

The proof of the lemma is by induction on the structure of $\psi$. The base case with propositional variables and the cases with Boolean operators in the induction step are by an easy verification.

*Case 1*: $\psi = \mathtt{X}^{\beta'}\varphi$ with $\beta' \geq 0$

By Lemma C.1, $\psi \in \rho(\beta)$ iff $\varphi \in \rho(\beta + \beta')$. By induction hypothesis, $\varphi \in \rho(\beta + \beta')$ is equivalent to $\sigma, \beta + \beta' \models \varphi$ which is equivalent to $\sigma, \beta \models \psi$ by definition of $\models$.

*Case 2*: $\psi = \psi_1 \mathtt{U}^{\beta'} \psi_2$

The propositions below are equivalent:

1. $\psi \in \rho(\beta)$,

2. there is $\beta \leq \gamma < \beta + \beta'$ such that $\psi_2 \in \rho(\gamma)$ and for every $\beta \leq \gamma' < \gamma$, $\psi_1 \in \rho(\gamma')$ (by (hin4))

3. there is $\beta \leq \gamma < \beta + \beta'$ such that $\sigma, \gamma \models \psi_2$ and for every $\beta \leq \gamma' < \gamma$, $\sigma, \gamma' \models \psi_1$ (by induction hypothesis)

4. $\sigma, \beta \models \psi$ (by definition of $\models$).

$\square$

## Appendix D: Proof of Lemma 7

We show that the set of sequences $\rho : \omega^k \to 2^{cl(\phi)}$ obtained from accepting runs $r : \omega^k + 1 \to Q$ of $\mathcal{A}_\phi$ as described below is precisely the set of Hintikka sequences for $\phi$. $\rho$ is defined from $r$ as follows: for every $\alpha < \omega^k + 1$, $\rho(\alpha) = X$ where $r(\alpha) = \langle X, tail(\alpha) \rangle$.

**(I)** First, we show that if $r : \omega^k + 1 \to Q$ is an accepting run, then $\rho$ is an Hintikka sequence for $\phi$. Satisfaction of (hin1) and (hin2) is immediate.

**(hin3)** We want to show that for all $\alpha < \omega^k$, $\mathtt{X}^{\alpha'}\psi \in cl(\phi)$ and $0 \leq n \leq k$ such that $\alpha' \geq \omega^n$, $\mathtt{X}^{\alpha'}\psi \in \rho(\alpha)$ iff $\mathtt{X}^{\alpha'-\omega^n}\psi \in \rho(\alpha + \omega^n)$. When $n = 0$, the property is satisfied thanks to (A3) in $\mathcal{A}_\phi$. Otherwise suppose that $\mathtt{X}^{\alpha'}\psi \in \rho(\alpha)$ and $\alpha' \geq \omega^n$ with $n \geq 1$. We can show by transfinite induction, that for every $\beta < \omega^n$, $\mathtt{X}^{\alpha'}\psi \in \rho(\alpha + \beta)$. The base case $\beta = 0$ is obvious. In the induction step with $\beta + 1 < \omega^n$, (A3) guarantees that $\mathtt{X}^{\alpha'}\psi \in \rho(\alpha + \beta)$ implies $\mathtt{X}^{\alpha'}\psi \in \rho(\alpha + \beta + 1)$ since $\alpha' - 1 = \alpha'$ (remember $\alpha' \geq \omega$). Now suppose $\beta$ is a limit ordinal strictly smaller than $\omega^n$ and for every $\beta' < \beta$, $\mathtt{X}^{\alpha'}\psi \in \rho(\alpha + \beta')$. By (A5), $\mathtt{X}^{\alpha'-\omega^{tail(\beta)}}\psi \in \rho(\alpha + \beta)$. Since $\beta < \omega^n$ and $\alpha' \geq \omega^n$, $\alpha' - \omega^{tail(\beta)} = \alpha'$. So, for every $\beta < \omega^n$, $\mathtt{X}^{\alpha'}\psi \in \rho(\alpha + \beta)$. By (A5), we obtain $\mathtt{X}^{\alpha'-\omega^n}\psi \in \rho(\alpha + \omega^n)$.

Now suppose that $\mathtt{X}^{\alpha'-\omega^n}\psi \in \rho(\alpha + \omega^n)$ with $\alpha' \geq \omega^n$ and $n \geq 1$. So there is a limit transition $Z \to \langle \rho(\alpha + \omega^n), n\rangle$ such that for every $\langle Y, n'\rangle \in Z$, $\mathtt{X}^{\alpha'}\psi \in Y$. Since $Z = inf(\alpha + \omega^n, r)$, there is $\alpha \leq \beta < \alpha + \omega^n$ such that $\mathtt{X}^{\alpha'}\psi \in \rho(\beta)$. We can now show that for every $\alpha \leq \beta' \leq \beta$, $\mathtt{X}^{\alpha'}\psi \in \rho(\beta')$. This can be proved as above by observing that for such $\beta'$, $tail(\beta') < n$ and therefore $\alpha' - \omega^{tail(\beta')} = \alpha'$.

**(hin4)** We show that for all $\alpha < \omega^k$ and $\psi_1 \mathtt{U}^{\alpha'}\psi_2 \in cl(\phi)$, (A) $\psi_1 \mathtt{U}^{\alpha'}\psi_2 \in \rho(\alpha)$ iff (B) there is $\alpha \leq \alpha'' < \alpha + \alpha'$ such that $\psi_2 \in \rho(\alpha'')$ and for every $\alpha \leq \beta < \alpha''$, $\psi_1 \in \rho(\beta)$.

If $sum(\alpha') = 0$ or $\alpha' = 1$, then the proof is immediate since $\rho(\alpha)$ satisfies (mc5) and (mc8), respectively.

The proof is by induction on $sum(\alpha')$ with obvious base case $sum(\alpha') = 0$.

*Base case 1*: $sum(\alpha') = 1$.
Suppose $\alpha' = \omega^N$ for some $1 \leq N \leq k$.

Proof of "(A) implies (B)".
The proof is by induction on $N$.
*Case 1*: $N = 1$ ("LTL case").
Suppose $\psi_1 \mathtt{U}^\omega \psi_2 \in \rho(\alpha)$. If $\psi_1 \mathtt{U}^1 \psi_2 \in \rho(\alpha)$, then (B) trivially holds. Otherwise, $\psi_1, \mathtt{X}^1(\psi_1 \mathtt{U}^\omega \psi_2) \in \rho(\alpha)$ and by (mc5) $\psi_1 \mathtt{U}^0 \psi_2 \notin \rho(\alpha + \omega)$. By definition of limit transitions, there is $\alpha \leq \alpha + i < \alpha + \omega$ such that $r(\alpha + i) \in P_{\psi_1 \mathtt{U}^\omega \psi_2}$ with $i \geq 0$. Take the minimal $i$ satisfying this property. So for every $0 \leq j < i$, $\psi_1 \mathtt{U}^\omega \psi_2, \neg \psi_2 \in \rho(\alpha + j)$.

Suppose that $\psi_1 \mathtt{U}^\omega \psi_2 \notin \rho(\alpha + i)$ and not $\psi_2 \in \rho(\alpha + i)$. If $i = 0$ this leads immediately to a contradiction. Otherwise, by (mc6) $\psi_1 \mathtt{U}^\omega \psi_2, \neg \psi_2 \in \rho(\alpha + (i-1))$ implies $\mathtt{X}^1(\psi_1\mathtt{U}^\omega\psi_2) \in \rho(\alpha+(i-1))$ which is in contradiction with $\psi_1\mathtt{U}^\omega\psi_2 \notin \rho(\alpha+i)$. So $\psi_2 \in \rho(\alpha+i)$. By (mc6), for every $0 \leq j < i$, $\psi_1\mathtt{U}^\omega\psi_2, \neg\psi_2 \in \rho(\alpha+j)$ implies $\psi_1 \in \rho(\alpha+j)$. Hence, for every $0 \leq j < i$, $\psi_1 \in \rho(\alpha+j)$ and $\psi_2 \in \rho(\alpha+i)$. So (B) holds true.

*Case 2*: $N > 1$.
Suppose $\psi_1\mathtt{U}^{\omega^N}\psi_2 \in \rho(\alpha)$. If there is $N' < N$ such that $\psi_1\mathtt{U}^{\omega^{N'}}\psi_2 \in \rho(\alpha)$, then by induction hypothesis, (B) holds true. Otherwise, let us treat the case when for every $N' < N$, $\psi_1\mathtt{U}^{\omega^{N'}}\psi_2 \notin \rho(\alpha)$. Since $\psi_1\mathtt{U}^0\psi_2 \notin \rho(\alpha+\omega^N)$, there is $\alpha^\star$ such that $\alpha \leq \alpha + \alpha^\star < \alpha + \omega^N$ and $r(\alpha+\alpha^\star) \cap P_{\psi_1\mathtt{U}^{\omega^N}\psi_2}$ is non-empty. Take $\alpha^\star$ to be the minimal such an ordinal. It exists since the set of ordinals is well-ordered.

*Case 2.1*: $\psi_2 \in \rho(\alpha+\alpha^\star)$.
By minimality, for every $\alpha \leq \beta < \alpha+\alpha^\star$, $\psi_1\mathtt{U}^{\omega^N}\psi_2, \neg\psi_2 \in \rho(\beta)$. By (mc6), for every $\alpha \leq \beta < \alpha+\alpha^\star$, $\psi_1 \in \rho(\beta)$. So (B) holds true.

*Case 2.2*: $\psi_2 \notin \rho(\alpha+\alpha^\star)$.
Consequently, $\psi_1\mathtt{U}^{\omega^N}\psi_2 \notin \rho(\alpha+\alpha^\star)$ since $r(\alpha+\alpha^\star) \cap P_{\psi_1\mathtt{U}^{\omega^N}\psi_2}$ $neq\emptyset$. We shall show that this case leads to a contradiction.

*Case 2.2.1*: $\alpha+\alpha^\star$ is a successor ordinal, say $\alpha^\star = \alpha_0^\star + 1$.
Since $\psi_1\mathtt{U}^{\omega^N}\psi_2, \neg\psi_2 \in \rho(\alpha+\alpha_0^\star)$ by minimality, satisfaction of (mc6) implies

$$\mathtt{X}^1(\psi_1\mathtt{U}^{\omega^N}\psi_2) \in \rho(\alpha+\alpha_0^\star).$$

Hence, $\psi_1\mathtt{U}^{\omega^N}\psi_2 \in \rho(\alpha+\alpha_0^\star+1)$, a contradiction.

*Case 2.2.2*: $\alpha+\alpha^\star$ is a limit ordinal.
There is a limit transition $Z \to r(\alpha+\alpha^\star)$ such that $inf(\alpha+\alpha^\star, r) = Z$. Since $\omega^N - \omega^{tail(\alpha^\star)} = \omega^N$ ($\alpha^\star < \omega^N$) and $\psi_1\mathtt{U}^{\omega^N}\psi_2 \notin \rho(\alpha+\alpha^\star)$, there is $\langle Y, n' \rangle \in Z$ such that $\langle Y, n' \rangle \in P_{\psi_1\mathtt{U}^{\omega^N}\psi_2}$. As $inf(\alpha+\alpha^\star, r) = Z$, there is $\alpha \leq \beta < \alpha+\alpha^\star$ such that $r(\beta) \in P_{\psi_1\mathtt{U}^{\omega^N}\psi_2}$, a contradiction by minimality of $\alpha+\alpha^\star$.

Proof of "(B) implies (A)" (in *Base case 1*).
We show a bit stronger property: for all $\alpha < \omega^k$, $\psi_1\mathtt{U}^{\alpha'}\psi_2 \in cl(\phi)$ and $0 \leq \alpha'' < \alpha'$, if $\psi_1\mathtt{U}^{\alpha'-\alpha''}\psi_2 \in \rho(\alpha+\alpha'')$ and for every $0 \leq \gamma < \alpha''$, $\psi_1 \in \rho(\alpha+\gamma)$, then $\psi_1\mathtt{U}^{\alpha'}\psi_2 \in \rho(\alpha)$. By Lemma 3(III), we know that $\psi_1\mathtt{U}^{\alpha'-\alpha''}\psi_2 \in cl(\phi)$. So if (B) holds true, that is, there is $0 \leq \alpha'' < \alpha'$ such that $\psi_2 \in \rho(\alpha+\alpha'')$ and for every $0 \leq \beta < \alpha''$, $\psi_1 \in \rho(\alpha+\beta)$, then $\psi_1\mathtt{U}^{\alpha'}\psi_2 \in \rho(\alpha)$. Indeed, by (mc8), $\psi_1\mathtt{U}^1\psi_2 \in \rho(\alpha+\alpha'')$ and by (mc7), $\psi_1\mathtt{U}^{\alpha'-\alpha''}\psi_2 \in \rho(\alpha+\alpha'')$.
The proof is by structural induction on $\alpha''$.

*Base case 2*: $\alpha'' = 0$.
Immediate.

*Induction step 2.*
We distinguish two cases depending whether $\alpha''$ is a limit ordinal or not.

*Case 1*: $\alpha'' = \alpha^\star + 1$.
Since $\alpha' - (\alpha^\star + 1) = (\alpha' - \alpha^\star) - 1$ and by hypothesis, $\psi_1 \in \rho(\alpha+\alpha^\star)$ and $\mathtt{X}^1(\psi_1\mathtt{U}^{(\alpha'-\alpha^\star)-1}\psi_2) \in \rho(\alpha+\alpha^\star)$. By (mc6), $\psi_1\mathtt{U}^{(\alpha'-\alpha^\star)}\psi_2 \in \rho(\alpha+\alpha^\star)$ and by induction hypothesis, $\psi_1\mathtt{U}^{\alpha'}\psi_2 \in \rho(\alpha)$.

*Case 2*: $\alpha''$ is a limit ordinal ($tail(\alpha'') \geq 1$).

Suppose that $\alpha'' = \alpha^\star + \omega^{tail(\alpha'')} \times n$ for some $n \geq 1$. There is a limit transition $Z \to \langle r(\alpha + \alpha''), tail(\alpha'') \rangle$ in $\mathcal{A}_\phi$ such that $inf(\alpha + \alpha'', r) = Z$. By (mc4), $\mathtt{X}^0(\psi_1 \mathtt{U}^{\alpha' - \alpha''} \psi_2) \in \rho(\alpha + \alpha'')$. By (A5), for every $\langle Y, n' \rangle \in Z$,

$$\mathtt{X}^{\omega^{tail(\alpha'')}}(\psi_1 \mathtt{U}^{\alpha' - \alpha''} \psi_2) \in Y.$$

Since $inf(\alpha + \alpha'', r) = Z$, there is $\beta \geq \alpha^\star + \omega^{tail(\alpha'')} \times (n - 1)$ such that for every $\beta \leq \gamma < \alpha''$, $r(\alpha + \gamma) \in Z$. Hence, for every $\beta \leq \gamma < \alpha''$, $\mathtt{X}^{\omega^{tail(\alpha'')}}(\psi_1 \mathtt{U}^{\alpha' - \alpha''} \psi_2) \in \rho(\alpha + \gamma)$. Since (A) implies (B) and by hypothesis, for every $\beta \leq \gamma < \alpha''$, $\psi_1 \in \rho(\alpha + \gamma)$, we have that for every $\beta \leq \gamma < \alpha''$, $\neg(\top \mathtt{U}^{\omega^{tail(\alpha'')}} \neg \psi_1) \in \rho(\alpha + \gamma)$. So by (mc6), $\psi_1 \mathtt{U}^{\omega^{tail(\alpha'')} + (\alpha' - \alpha'')} \psi_2 \in \rho(\alpha + \gamma)$ for every $\beta \leq \gamma < \alpha''$. Since for every $\beta \leq \gamma < \alpha''$, $\gamma + \omega^{tail(\alpha'')} = \alpha''$, we have $\omega^{tail(\alpha'')} + (\alpha' - \alpha'') = \alpha' - \gamma$. Hence, for every $\beta \leq \gamma < \alpha''$, $\psi_1 \mathtt{U}^{\alpha' - \gamma} \psi_2 \in \rho(\alpha + \gamma)$. In particular, $\psi_1 \mathtt{U}^{\alpha' - \beta} \psi_2 \in \rho(\alpha + \beta)$. By induction hypothesis, $\psi_1 \mathtt{U}^{\alpha'} \psi_2 \in \rho(\alpha)$.

*Induction step 1*: $sum(\alpha') > 2$.
By (mc6), $\psi_1 \mathtt{U}^{\alpha'} \psi_2 \in \rho(\alpha)$ iff

- either $\psi_1 \mathtt{U}^{\omega^N} \psi_2 \in \rho(\alpha)$,
- or $\neg(\top \mathtt{U}^{\omega^N} \neg \psi_1), \mathtt{X}^{\omega^N}(\psi_1 \mathtt{U}^{\alpha' - \omega^N} \psi_2) \in \rho(\alpha)$

where $N = head(\alpha')$. Since $sum(\omega^N) < sum(\alpha')$ and $sum(\alpha' - \omega^N) < sum(\alpha')$, by induction hypothesis we obtain that either there is $\alpha \leq \alpha'' < \alpha + \omega^N$ such that $\psi_2 \in \rho(\alpha'')$ and for every $\alpha \leq \beta < \alpha''$, $\psi_1 \in \rho(\beta)$ or for every $\alpha \leq \beta < \alpha + \omega^N$, $\psi_1 \in \rho(\beta)$ and $\mathtt{X}^{\omega^N}(\psi_1 \mathtt{U}^{\alpha' - \omega^N} \psi_2) \in \rho(\alpha)$. Since $\rho$ satisfies (hin3), $\mathtt{X}^{\omega^N}(\psi_1 \mathtt{U}^{\alpha' - \omega^N} \psi_2) \in \rho(\alpha)$ iff $(\psi_1 \mathtt{U}^{\alpha' - \omega^N} \psi_2) \in \rho(\alpha + \omega^N)$. By induction hypothesis, we obtain that for every $\alpha \leq \beta < \alpha + \omega^N$, $\psi_1 \in \rho(\beta)$ and $\mathtt{X}^{\omega^N}(\psi_1 \mathtt{U}^{\alpha' - \omega^N} \psi_2) \in \rho(\alpha)$ is equivalent to: for every $\alpha \leq \beta < \alpha + \omega^N$, $\psi_1 \in \rho(\beta)$ and there is $\alpha + \omega^N \leq \alpha'' < \alpha + \alpha'$ such that $\psi_2 \in \rho(\alpha'')$ and for every $\alpha + \omega^N \leq \beta < \alpha''$, $\psi_1 \in \rho(\beta)$. Hence, (B) holds true.

**(II)** Now we show that for every Hintikka sequence $\rho$ for $\phi$, the sequence $r : \omega^k + 1 \to Q$ defined by $r(\alpha) = \langle \rho(\alpha), tail(\alpha) \rangle$ is an accepting run of $\mathcal{A}_\phi$. For technical reason, suppose also that $r(\omega^k + 1)$ takes an arbitrary value of the form $\langle X, k \rangle$. Observe that

- $\langle \rho(0), 0 \rangle \in I$ since $\phi \in \rho(0)$ by (hin1),
- $\langle \rho(\omega^k), k \rangle \in F$,
- for every $0 \leq \alpha < \omega^k$, $\langle \rho(\alpha), tail(\alpha) \rangle \to \langle \rho(\alpha + 1), 0 \rangle$ since $\rho$ satisfies (hin3).

The only property that really deserves to be checked is that for every limit ordinal $\alpha$, $inf(\alpha, r) \to \langle \rho(\alpha), tail(\alpha) \rangle$ is a valid limit transition of $\mathcal{A}_\phi$. We write $\alpha = \alpha^\star + \omega^{tail(\alpha)} \times n$ for some $n \geq 1$. Since $\alpha$ is a limit ordinal, we also have $tail(\alpha) \geq 1$: condition (A6) is satisfied.
Observe that

$$inf(\alpha, r) \subseteq \{r(\beta) : \alpha^\star + \omega^{tail(\alpha)} \times (n - 1) < \beta < \alpha\}.$$

So for every $\langle Y, n' \rangle \in inf(\alpha, r)$, $n' \leq n-1$: condition (A4) is satisfied.
Since
$$\{r(\alpha^\star + \omega^{tail(\alpha)} \times (n-1) + \omega^{tail(\alpha)-1} \times i) : i \geq 0\}$$
is finite, there is $m \geq 0$ such that $r(\alpha^\star + \omega^{tail(\alpha)} \times (n-1) + \omega^{tail(\alpha)-1} \times m) \in inf(\alpha, r)$: condition (A8) is therefore satisfied.

Let $\beta$ be such that $\alpha^\star + \omega^{tail(\alpha)} \times (n-1) < \beta$ and $r(\beta) \in inf(\alpha, r)$. We have $\beta + \omega^{tail(\alpha)} = \alpha$ and by satisfaction of (hin3), for all $\alpha' \geq \omega^{tail(\alpha)}$ and $\mathtt{X}^{\alpha'}\psi \in cl(\phi)$, $\mathtt{X}^{\alpha'}\psi \in \rho(\beta)$ iff $\mathtt{X}^{\alpha'-\omega^{tail(\alpha)}}\psi \in \rho(\alpha)$. Hence, condition (A5) is satisfied.

It remains to check that condition (A9) holds true. Let $\beta$ be such that $\alpha^\star + \omega^{tail(\alpha)} \times (n-1) < \beta$ and $r(\beta) \in inf(\alpha, r)$, $\alpha'$ be such that $\alpha' \geq \omega^{tail(\alpha)}$ and $\psi_1 \mathtt{U}^{\alpha'} \psi_2 \in cl(\phi)$. Since $r(\beta) \in inf(\alpha, r)$, there is a countable family of ordinals $(\beta_i)_{i \in \mathbb{N}}$ such that

- $\beta_0 = \beta$,
- for every $i \geq 0$,
  - $r(\beta_i) = r(\beta)$,
  - $\beta_i < \beta_{i+1} < \alpha$.
- for every $\gamma$ such that $\alpha^\star + \omega^{tail(\alpha)} \times (n-1) < \gamma < \alpha$, there is $j \geq 0$ such that $\gamma < \beta_j$.

By satisfaction of (mc6), $\psi_1 \mathtt{U}^{\alpha'} \psi_2 \in \rho(\beta)$ iff either

$$\psi_1 \mathtt{U}^{\omega^{tail(\alpha)}} \psi_2 \in \rho(\beta), \text{ or}$$

$$\neg(\top \mathtt{U}^{\omega^{tail(\alpha)}} \neg \psi_1), \mathtt{X}^{\omega^{tail(\alpha)}}(\psi_1 \mathtt{U}^{\alpha'-\omega^{tail(\alpha)}} \psi_2) \in \rho(\beta).$$

Suppose that $\psi_1 \mathtt{U}^{\alpha'-\omega^{tail(\alpha)}} \psi_2 \notin \rho(\alpha)$. By satisfaction of (A5), we get

$$\mathtt{X}^{\omega^{tail(\alpha)}}(\psi_1 \mathtt{U}^{\alpha'-\omega^{tail(\alpha)}} \psi_2) \notin \rho(\beta).$$

So $\psi_1 \mathtt{U}^{\alpha'} \psi_2 \in \rho(\beta)$ iff $\psi_1 \mathtt{U}^{\omega^{tail(\alpha)}} \psi_2 \in \rho(\beta)$. If $\psi_1 \mathtt{U}^{\alpha'} \psi_2 \notin \rho(\beta)$ then $r(\beta) \in P_{\psi_1 \mathtt{U}^{\alpha'} \psi_2}$. Otherwise, for every $i \geq 0$, $\psi_1 \mathtt{U}^{\alpha'} \psi_2 \in \rho(\beta_i)$ and therefore for every $i \geq 0$, $\psi_1 \mathtt{U}^{\omega^{tail(\alpha)}} \psi_2 \in \rho(\beta_i)$. Since for every $i \geq 0$, $\beta_i + \omega^{tail(\alpha)} = \alpha$, by satisfaction of (hin4), we obtain that for every $i \geq 0$, there is $\beta_i \leq \beta_i' < \alpha$ such that $\psi_2 \in \rho(\beta_i')$. So there is a family of ordinals $(\beta_{t_i}')_{i \in \mathbb{N}}$ such that

- $\beta_{t_0}' = \beta_l'$ for some $l \geq 0$,
- for every $i \geq 0$,
  - $r(\beta_{t_i}') = r(\beta_l')$,
  - $\beta_{t_i}' < \beta_{t_{i+1}}' < \alpha$.
- for every $\gamma$ such that $\alpha^\star + \omega^{tail(\alpha)} \times (n-1) < \gamma < \alpha$, there is $j \geq 0$ such that $\gamma < \beta_{t_j}'$.

Consequently, $r(\beta'_l) \in inf(\alpha, r)$ and $\psi_2 \in r(\beta'_l)$, which means that $r(\beta'_l) \in P_{\psi_1 \mathtt{U}^{\alpha'} \psi_2}$. Hence, condition (A9) is satisfied. □

**Appendix E: Proof of Lemma 8**

The EXPSPACE upper bound can be obtained by designing an obvious exponential space translation into LTL which is known to be PSPACE-complete [42]. Indeed,

1. $\mathtt{X}^n \phi$ is equivalent to $\overbrace{\mathtt{X} \cdots \mathtt{X}}^{n \text{ times}} \phi$,

2. $\phi_1 \mathtt{U}^\omega \phi_2$ is equivalent to $\phi_1 \mathtt{U} \phi_2$, and

3. $\phi_1 \mathtt{U}^n \phi_2$ ($n \geq 1$) is equivalent $\phi_2 \vee (\phi_1 \wedge \mathtt{X}(\phi_1 \mathtt{U}^{n-1} \phi_2))$ and $\phi_1 \mathtt{U}^0 \phi_2$ is equivalent to $\bot$.

In order to show the EXPSPACE lower bound, we present a reduction from the $2^n$-corridor tiling problem that is EXPSPACE-complete, see [44] and references therein. A tile is a unit square of one of the several tile-types and the tiling problem we considered is specified by means of a finite set $T$ of tile-type (say $T = \{t_1, \ldots, t_l\}$), two binary relations $H$ and $V$ over $T$ and two distinguished tile-types $t_{init}, t_{final} \in T$. The tiling problem consists in determining whether, for a given number $n$ in unary, the region $[0, \ldots, 2^n - 1] \times [0, \ldots, k - 1]$ of the integer plane for some $k$ can be tiled consistently with $H$ and $V$, $t_{init}$ is the left bottom tile, and $t_{final}$ is the right upper tile.

Given an instance $I = \langle T, t_{init}, t_{final}, n \rangle$ of the tiling problem, we build a formula $\phi_I$ such that $I = \langle T, t_{init}, t_{final}, n \rangle$ has a solution iff $\phi_I$ is LTL($\omega$) satisfiable. For $t \in T$, we introduce the propositional variable $p_t$. Additionally, we introduce the variable $p_{end}$ stating that the end of the tiling plane is reached and $p_{newline}$ stating that a new line starts. The formula $\phi_I$ is the conjunction of the following formulae:

- The region of the integer plane for the solution is finite:

$$\neg p_{end} \wedge (\neg p_{end} \mathtt{U}^\omega (p_{newline} \wedge \mathtt{G}^\omega p_{end})).$$

- There is exactly one tile per element of the plane region:

$$\mathtt{G}^\omega (\neg p_{end} \Rightarrow \bigvee_{t \in T} (p_t \wedge \bigwedge_{t' \neq t} \neg p_{t'})).$$

- Constraint on the right upper tile:

$$\mathtt{F}^\omega (p_{t_{final}} \wedge \mathtt{X}^1 p_{end}).$$

- Constraint on the left bottom tile:

$$p_{newline} \wedge p_{t_{init}}.$$

- New line:

$$\mathtt{G}^\omega (p_{newline} \Leftrightarrow \mathtt{X}^{2^n} p_{newline}) \wedge \mathtt{X}^1 \neg (\top \mathtt{U}^{2^n - 1} p_{newline})$$

- Horizontal consistency:

$$\mathtt{G}^\omega((\overbrace{(\neg \mathtt{X}^1 p_{newline})}^{\text{not the last element of a row}} \wedge \neg p_{end}) \Rightarrow \bigwedge_{t \in T}(p_t \Rightarrow \bigvee_{\langle t, t_{hor}\rangle \in H} \mathtt{X}^1 p_{t_{hor}})).$$

- Vertical consistency:

$$\mathtt{G}^\omega(\bigwedge_{t \in T}(p_t \wedge \overbrace{\neg p_{end} \wedge \mathtt{F}^\omega(\mathtt{X}^1 \neg p_{end} \wedge \mathtt{X}^1 p_{newline})}^{\text{not on the last row}}) \Rightarrow \bigvee_{\langle t, t_{ver}\rangle \in V} \mathtt{X}^{2^n} p_{t_{ver}}).$$

One can show that the instance $I = \langle T, t_{init}, t_{final}, n\rangle$ has a solution iff $\phi_I$ is LTL$(\omega)$ satisfiable. □

**Appendix F: Proof of Lemma 12**

Let $r$ be a run of length $\omega^{k'}$ ($r(0) \in I$). There is a rule $P \to q' \in E$ such that

- $l(q') = k'$ and $r(\omega^{k'}) = q'$ (by Lemma 10),
- $inf(\omega^{k'}, r) = P$,
- there is $n > 0$ such that $P = \{q \in Q : \beta \geq \omega^{k'-1} \times n, \ r(\beta) = q\}$.

$\omega^{k'-1} \times n$ is the ordinal after which all states $r(\beta)$ with $\beta \geq \omega^{k'-1} \times n$ occurs infinitely often. Suppose there exist $n_1, n_2 \geq 0$ such that $n_1 < n_2 < n$ and $r(\omega^{k'-1} \times n_1) = r(\omega^{k'-1} \times n_2)$. Then $r'$ defined below is also a run of length $\omega^{k'}$ with $r'(0) \in I$:

- for every $\beta < \omega^{k'-1} \times n_1$, $r'(\beta) \stackrel{\text{def}}{=} r(\beta)$.
- for every $\beta \geq \omega^{k'-1} \times n_1$ such that its normal form is $\omega^{k'-1} \times n'_1 + \gamma$, $r'(\beta) \stackrel{\text{def}}{=} r(\omega^{k'-1} \times (n'_1 + (n_2 - n_1)) + \gamma)$. We still have $P = \{q \in Q : \beta \geq \omega^{k'-1} \times (n - (n_2 - n_1)), \ r'(\beta) = q\}$.

By applying this transformation an adequate number of times (at most $n$ times), we can assume that $n \leq |Q|$ and we fix $K = n$.

Now we shall define $K'$. Assume that $P = \{q_0, \ldots, q_s\}$. We order the members of $P$ by decreasing level and the states with identical level are arbitrarily ordered. We introduce for technical reasons an artificial state $q_{s+1}$ equal to $q_0$. Without any loss of generality, we can assume that $r(\omega^{k'-1} \times n) = q_0$. We define

- a family $(n_i)_{0 \leq i \leq s+1}$ of natural numbers such that $0 \leq n_{i+1} - n_i \leq |Q|$,
- families $(\beta_i)_{0 \leq i \leq s+1}$ and $(\beta'_i)_{0 \leq i \leq s+1}$ of ordinals smaller than $\omega^{k'}$ such that the normal form of $\beta_i$ is $\omega^{k'-1} \times n_i + \beta'_i$.

The base case $i = 0$ is defined as follows: $\beta_0 = \omega^{k'-1} \times n$, $n_0 = n$ and $\beta'_0 = 0$. Then, let us define $n_{i+1}$, $\beta_{i+1}$ and $\beta'_{i+1}$ assuming that $n_i$, $\beta_i$ and $\beta'_i$ are already defined. Since $inf(\omega^{k'}, r) = P$ and $q_{i+1} \in P$, there is $\beta_{i+1} \geq \omega^{k'-1} \times n_i$ such that $r(\beta_{i+1}) = q_{i+1}$ and $\beta_{i+1} = \omega^{k'-1} \times n_{i+1} + \beta'_{i+1}$. By applying a reasoning similar to the one showing that $K \leq |Q|$, we can assume that $n_{i+1} - n_i \leq |Q|$.

Consequently, there are families $(n_i)_{0 \leq i \leq s+1}$, $(\beta_i)_{0 \leq i \leq s+1}$ and $(\beta'_i)_{0 \leq i \leq s+1}$ such that for every $i \in \{0, \ldots, s+1\}$

- $\beta_i = \omega^{k'-1} \times n_i + \beta'_i$,
- $r(\beta_i) = q_i$,
- $0 \leq n_{i+1} - n_i \leq |Q|$ for every $0 \leq i \leq s$.

Hence $n_{s+1} - n_0 \leq |Q|^2$. We fix $K' = n_{s+1} - n_0$. Then $r'$ defined below is also a run of length $\omega^{k'}$ ($r(0) \in I$):

- for every $\beta < \omega^{k'-1} \times (K + K')$, $r'(\beta) \stackrel{\text{def}}{=} r(\beta)$.
- for every $\beta \geq \omega^{k'-1} \times (K + K')$ such that its normal form is $\omega^{k'-1} \times n' + \gamma$, $r'(\beta) \stackrel{\text{def}}{=} r(\omega^{k'-1} \times (K+m) + \gamma)$ with $n' - K \equiv_{K'} m$ for some $0 \leq m \leq K' - 1$.

Observe that $P = \{q : K \leq \beta'' \leq K + K', r'(\beta'') = q\}$. Hence $r'$ satisfies the properties stated in Lemma 12. □

## Appendix G: Proof of Theorem 3

In the unary case, the PSPACE lower bound is a consequence of the PSPACE-hardness of LTL [42] whereas in the binary, EXPSPACE-hardness is a corollary of Lemma 8.

As far as the upper bound is concerned, in the unary [resp. binary] case, $\mathcal{A}'_\phi$ is of size exponential [resp. doubly exponential] in $|\phi|$ and requires only polynomial [resp. exponential] space to be built. By adapting the proof of [6, Corollary 3.36] and by considering Corollary 2 and Lemma 7, we obtain that given $\phi$, testing the emptiness of $L(\mathcal{A}_\phi)$ can be done in PSPACE [resp. EXPSPACE]. □

## Appendix H: Proof of Theorem 4

We show EXPSPACE-completeness with the binary encoding, the proof of the PSPACE-completeness with unary encoding being quite similar.

In order to establish, EXPSPACE-hardness, let us consider a deterministic Turing machine $M = \langle \Sigma, Q, q_0, \delta \rangle$ with transition function $\delta : Q \times \Sigma \to Q \times \Sigma \times \{-1, 0, 1\}$. $Q$ contains the special states for acceptance (called `accept` here), rejection and halting. Similarly, we assume that the alphabet $\Sigma$ contains the blank symbol `blank` and the left marker $\triangleright$. Finally, we assume that once the machine enters in the acceptance state, it loops on it without moving the read/write head and without changing the tape content. In order to show EXPSPACE-hardness, we suppose that $M$ runs in space $2^{n^K}$ with $n$ the size of the input for some $K \geq 1$.

Let $\Sigma'$ be the new alphabet $\Sigma \times (Q \times \Sigma)$ for the automaton $\mathcal{A}$ below:

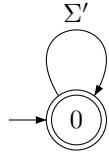

with unique limit transition $\{0\} \to 0$. The ordinal automaton $\mathcal{A}$ recognizes all the $\alpha$-sequences $\alpha \to \Sigma'$ and in particular all the sequences $\omega^k \to \Sigma$.

Let $x = x_1, \ldots, x_n$ be an input over the alphabet $\Sigma \setminus \{\triangleright, \texttt{blank}\}$. The conjunction $\phi$ of the formulae below encodes the existence of an accepting run of $M$ on input $x$ so that $\mathcal{A} \models \phi$ iff $M$ has an accepting run on input $x$. The formula $\phi$ is the conjunction of the formulae below.

- Input word is $x$:

$$\triangleright \wedge \mathtt{X}^1\langle q_0, x_1\rangle \wedge \mathtt{X}^2 x_2 \wedge \ldots \wedge \mathtt{X}^n x_n \wedge \mathtt{X}^n \mathtt{G}^{2^{n^K}-n}\texttt{blank}.$$

- Reaching accepting configuration:

$$\mathtt{F}^\omega(\bigvee_{a \in \Sigma} \langle \texttt{accept}, a\rangle)$$

- Updating configuration ("the head is far away"):

$$\mathtt{G}^\omega(\bigwedge_{a,b,c \in \Sigma} (a \wedge \mathtt{X}^1 b \wedge \mathtt{X}^2 c) \Rightarrow \mathtt{X}^{2^{n^K}+1} b).$$

- Move of the head to the right:

$$\mathtt{G}^\omega(\bigwedge_{a,b,c,q,\delta(q,b)=\langle q',b',1\rangle} (a \wedge \mathtt{X}^1\langle q,b\rangle \wedge \mathtt{X}^2 c) \Rightarrow \mathtt{X}^{2^{n^K}} a \wedge \mathtt{X}^{2^{n^K}+1} b' \wedge \mathtt{X}^{2^{n^K}+2}\langle q',b'\rangle).$$

- Similar formulae for the move to the left and for no move.

Now, let us show that model-checking for $\mathrm{LTL}(\omega^k)$ is in EXPSPACE by reducing model-checking to satisfiability in logarithmic space. Let $\mathcal{A} = \langle Q, \Sigma, \delta, E, I, F\rangle$ be an ordinal automaton and $\phi$ be an $\mathrm{LTL}(\omega^k)$ formula such that $\Sigma$ is a subset of $2^{\mathrm{AP}(\phi)}$ and $\mathrm{AP}(\phi)$ is the set of propositional variables occurring in $\phi$. Let $\mathcal{A}' = \langle Q', \Sigma', \delta', E', I', F'\rangle$ be the ordinal automaton below:

- $Q' = Q \times \{0, \ldots, k\}$, $\Sigma' = \Sigma$,
- $I' = I \times \{0\}$, $F' = F \times \{k\}$,
- $\langle q, i\rangle \xrightarrow{a} \langle q', i'\rangle \in \delta' \stackrel{\mathrm{def}}{\Leftrightarrow} q \xrightarrow{a} q' \in \delta$, $i' = 0$,
- $P \to \langle q, i\rangle \in E' \stackrel{\mathrm{def}}{\Leftrightarrow} i > 0$, $\max\{j : \langle q', j\rangle \in P\} = i - 1$ and $\{q' : \langle q', j\rangle \in P\} \to q \in E$.

This stratification of the states in $\mathcal{A}'$ guarantees that $L(\mathcal{A}') = L(\mathcal{A}) \cap \Sigma^{\omega^k}$. Observe that $\mathrm{card}(E')$ is less than $\mathrm{card}(E) \times \mathrm{card}(Q)^k$ which is still polynomial ($k$ is fixed). Hence, $\mathcal{A}'$ can be computed in logarithmic space in $|\mathcal{A}|$. Now let us encode the accepting runs of $\mathcal{A}'$ by an $\mathrm{LTL}(\omega^k)$ formula $\phi_{\mathcal{A}'}$ that is a conjunction of the formulae below over the propositional variables in $\mathrm{AP}(\phi) \cup Q'$. For $a \in \Sigma$, by the formula $a$ we mean $\bigwedge_{p \in a} p \wedge \bigwedge_{p \in (\mathrm{AP}(\phi) \setminus a)} \neg p$.

- Initial state:
$$\bigvee_{q \in I'} q$$

- Final state:
$$\bigvee_{P \to q \in E', q \in F'} (\bigwedge_{q' \in P} \mathtt{G}^{\omega^k} \mathtt{F}^{\omega^k} q') \wedge (\bigwedge_{q' \in (Q' \setminus P)} \neg \mathtt{G}^{\omega^k} \mathtt{F}^{\omega^k} q')$$

- Any position is labelled by a unique state:
$$\mathtt{G}^{\omega^k} \bigvee_{q \in Q'} (q \wedge \bigwedge_{q' \neq q} \neg q')$$

- Any position is labelled by a letter in $\Sigma$:
$$\mathtt{G}^{\omega^k} \bigvee_{a \in \Sigma} a$$

- One-step transitions:
$$\mathtt{G}^{\omega^k} (\bigwedge_{q \in Q', a \in \Sigma} q \wedge a \Rightarrow \bigvee_{q \xrightarrow{a} q' \in \delta'} \mathtt{X}^1 q')$$

- For each set of states $P$ with limit transitions $P \to \langle q_1, i \rangle, \ldots, P \to \langle q_N, i \rangle$ and $i > 0$, we have:
$$\mathtt{G}^{\omega^k}((\bigwedge_{q' \in P} \mathtt{G}^{\omega^i} \mathtt{F}^{\omega^i} q') \wedge (\bigwedge_{q' \in (Q' \setminus P)} \neg \mathtt{G}^{\omega^i} \mathtt{F}^{\omega^i} q') \Rightarrow \mathtt{X}^{\omega^i}(q_1 \vee \ldots \vee q_N))$$

We have $\mathcal{A} \models \phi$ iff $\phi_{\mathcal{A}'} \wedge \phi$ is $\mathrm{LTL}(\omega^k)$ satisfiable, whence the EXPSPACE upper bound. □

**Appendix I: Proof of Theorem 5**

In order to establish Theorem 5, we first show the lemma below.

**Lemma I.1** *The class of languages recognized by $x$-succinct ordinal automata of level $k$ is closed under intersection.*

**Proof.**

Let $\mathcal{A}_1 = \langle Q_1, \Sigma, \delta_1, E_1, I_1, F_1, l_1 \rangle$ and $\mathcal{A}' = \langle Q_2, \Sigma, \delta_2, E_2, I_2, F_2, l_2 \rangle$ be $x$-succinct ordinal automata of level $k$ over the alphabet $\Sigma$. We define the intersection $x$-succinct ordinal automaton $\mathcal{A} = \langle Q, \Sigma, \delta, E, I, F, l \rangle$ as follows:

- $Q = Q_1 \times Q_2$, $I = I_1 \times I_2$, $F = F_1 \times F_2$.
- $\langle q_1, q_2 \rangle \xrightarrow{a} \langle q'_1, q'_2 \rangle \in \delta$ iff $q_1 \xrightarrow{a} q'_1 \in \delta_1$, and $q_2 \xrightarrow{a} q'_2 \in \delta_2$.
- For every $\langle P_0, P_1, \ldots, P_n, q \rangle \in E_1$ and $\langle R_0, R_1, \ldots, R_m, q \rangle \in E_2$ such that $l_1(q) = l_2(q')$, $\langle P'_0, P'_1, \ldots, P'_n, R'_0, R'_1, \ldots, R'_m, \langle q, q' \rangle \rangle \in E$ with
  - $P'_i = \{\langle r, r' \rangle \in Q : r \in P_i, r' \in Q_2, \ l_1(r) = l_2(r')\}$,
  - $R'_i = \{\langle r, r' \rangle \in Q : r' \in R_i, r \in Q_1, \ l_1(r) = l_2(r')\}$.

Observe that $n + m \leq |Q|$.

The automaton $\mathcal{A}$ can be viewed as the synchronized product between $\mathcal{A}_1$ and $\mathcal{A}_2$ and $\mathrm{L}(\mathcal{A}) = \mathrm{L}(\mathcal{A}_1) \cap \mathrm{L}(\mathcal{A}_2)$. □

Hardness is by an easy verification from the complexity of the standard complexity results for the model checking fo $\mathrm{LTL}(\omega^k)$.

As far as the upper bound is concerned, in the unary [resp. binary] case, $\mathcal{A} \times \mathcal{A}'_\phi$ is of size exponential [resp. doubly exponential] in $|\phi|$ and requires only polynomial [resp. exponential] space to be built, see Lemma I.1. By adapting the proof of [6, Corollary 3.36] and by considering Corollary 2 and Lemma 7, we obtain that given an $x$-succinct ordinal automata $\mathcal{A}$ of level $k$ and a formula $\phi$ in $\mathrm{LTL}(\omega^k)$, testing the emptiness of $\mathcal{A} \times \mathcal{A}'_\phi$ can be done in PSPACE [resp. EXPSPACE]. □

**Appendix J: Proof of Proposition 5**

The proof is by an easy verification by observing that $q \xrightarrow{a} q'$ in $\mathcal{A}$ iff there is a path $r : \omega^{k-1} + 1 \to Q$ in $lift_k(\mathcal{A})$ such that $r(0) = \langle k-1, q \rangle$, $r(1) = \langle 0, q' \rangle$ and $r(\omega^{k-1}) = \langle k-1, q' \rangle$. □